\documentclass[11pt]{article}
\usepackage[a4paper,left=2.0cm,right=2.0cm,top=2.0cm,bottom=2.0cm]{geometry}
\usepackage[english]{babel}
\usepackage[nottoc]{tocbibind}
\usepackage{fancyhdr}
\usepackage{fancyhdr, graphicx}
\usepackage{mathtools}
\usepackage{amsmath}
\usepackage{bm}
\usepackage{systeme}
\usepackage{graphicx}
\usepackage{subcaption}
\usepackage{array}
\usepackage[dvipsnames]{xcolor}
\usepackage{standalone}
\usepackage{multirow}
\usepackage{amsfonts}
\usepackage{amssymb}
\usepackage{CJKutf8}
\usepackage{rotating}
\usepackage{float}% bind figure
\usepackage{algorithm}
\usepackage{algpseudocode}

\usepackage[shortlabels]{enumitem}
\usepackage{import}
\usepackage{titlesec}
\newcommand{\secfnt}{\fontsize{11}{17}}
\newcommand{\ssecfnt}{\fontsize{11}{14}}
\titleformat{\section}
{\normalfont\secfnt\bfseries}{\thesection}{1em}{}
\titleformat{\subsection}
{\normalfont\ssecfnt\bfseries}{\thesubsection}{1em}{}

\usepackage[flushleft]{threeparttable}
\usepackage[superscript,biblabel]{cite}
\usepackage[left]{lineno}
\usepackage{hyperref}
\hypersetup{
 	colorlinks = true,
 	linkcolor = blue,
 	anchorcolor = blue,
 	citecolor = blue,
 	filecolor = blue,
 	pdftitle={SBoTFlow: A Scalable framework using lattice Boltzmann method and Topology-confined mesh refinement for moving-body applications},
 	}
\date{}
\title{\large \textbf{SBoTFlow: A Scalable framework using lattice Boltzmann method and Topology-confined mesh refinement for moving-body applications}}
\author{Duc V. Nguyen$^{1}$, Dung V. Duong$^{1*}$\\
 $^{1}$School of Aerospace Engineering, University of Engineering and Technology,\\
 Vietnam National University, Ha Noi City, Vietnam\\
}
 
\usepackage{amsmath}
\let\Gamma\varGamma
\usepackage{mathtools}

\fancypagestyle{plain}{%
   \fancyhf{}
   \fancyfoot[C]{\small $^*$Corresponding authors: duongdv@vnu.edu.vn}
   
}
 
\begin{document}
\sloppy
\pagenumbering{arabic}
\maketitle
 
\section*{Abstract}
This paper proposes a scalable lattice-Boltzmann computational framework (SBoTFlow) for simulations of flexible moving objects in an incompressible fluid flow. Behavior of fluid flow formed from moving boundaries of flexible-object motions is obtained through the multidirect forcing immersed boundary scheme associated with the lattice Boltzmann equation with a parallel topology-confined block refinement framework. We first demonstrate that the hydrodynamic quantities computed in this manner for standard benchmarks, including the Tayler-Green vortex flow and flow over an obstacle-embedded lid-driven cavity and an isolated circular cylinder, agree well with those previously published in the literature. We then exploit the framework to probe the underlying dynamic properties contributing to fluid flow under flexible motions at different Reynolds numbers by simulating large-scale flapping wing motions of both amplitude and frequency. The analysis shows that the proposed numerical framework for pitching and flapping motions has a strong ability to accurately capture high amplitudes, specifically up to $64^\circ$, and a frequency of $f=1/2.5\pi$. This suggests that the present parallel numerical framework has the potential to be used in studying flexible motions, such as insect flight or wing aerodynamics.

\textit{Key words}: Lattice Boltzmann method, immersed boundary method, flapping wing.

\section{Introduction}
Moving boundaries of complex geometries are regularly observed in engineering and nature applications, such as insect flight\cite{sane2003aerodynamics, nakata2012fluid}, rotorcraft\cite{abdelmaksoud2020control}, micro aerial vehicles (MAV) and automotive underwater vehicles (AUV)\cite{suzuki2019effect, huang2019recent, majumdar2020capturing}, thermal diffusion \cite{tao2022lattice}, and flow control\cite{greenblatt2022flow, xiao2013flow}. In order to simulate these applications, finite-volume (FVM) \cite{kim2001immersed, wang2015immersed} and finite-difference (FDM) \cite{kim2019immersed} methods are frequently used, while moving boundaries can be modeled by ghost-fluid \cite{tseng2003ghost, terashima2009front} and immersed boundary methods (IBM) \cite{peskin1972flow, peskin1977numerical}. The ghost-fluid method is for simple geometries, whereas the IBM method is for complex geometries. For an efficient fluid solver, computing speed and accuracy are two primary criteria. However, FVM and FDM-based solvers require a large computational cost to meet the convergence criterion of the Poisson equation \cite{nishiguchi2019full}. Otherwise, the lattice Boltzmann method (LBM) is increasingly attracting scientists' attention due to its second-order accuracy and straightforward parallelization. \cite{kruger2017lattice}. In this paper, we develop a robust and efficient incompressible fluid solver based on a combination of LBM and IBM for moving boundaries.

In the last two decades, the Lattice-Boltzmann method (LBM) has grown rapidly and gained popularity in computational fluid dynamics (CFD). LBM was initially adopted from the lattice gas automata method \cite{chen1998lattice}. Therefore, LBM has become a robust alternative to conventional CFD methods due to its simple algorithm and ability to simplify complex boundaries \cite{yeomans2006mesoscale}. On the other hand, LBM has also emerged as a parallel and efficient approach for simulating single- and multiphase fluid flows \cite{lu2022analyses, wang2023recent}. In the LBM approach, the particle distribution function is used to solve the kinetic equation; this avoids the need to solve the Poisson equation in conventional CFD approaches, leading to easy parallelism. Macroscopic variables (velocity and pressure) are obtained using the moments of the particle distribution function \cite{he1997priori, kruger2017lattice}. The main operator of LBM is the collision model, which is used to handle the Lattice Boltzmann equation \cite{succi2018lattice}; A popular and simple model in the LBM community is the Bhatnagar-Gross-Krook (BGK) collision model \cite{bhatnagar1954model}. By Chapman–Enskog analysis, LBM can recover the continuity and momentum equations for incompressible flow under the low Mach number limit \cite{he1997theory}. 

In order to model the stationary and moving boundaries, the IBM is often considered because of its simplicity. IBM was first introduced by Peskin\cite{peskin1977numerical} to study heart valve performance in 1977; it has been expanded to various scientific and engineering applications\cite{ou2022directional, liu2023investigation}. The IBM strategy is to mimic the interaction between the fluid and obstacle structures by directly adding an additional boundary force to the Navier-Stokes equations\cite{peskin1996case, peskin2002immersed}. The boundaries of solid geometries are naturally represented by a population of Lagrangian points immersed in Eulerian grids. Interpolation stencils are used to exchange the boundary force in communication between the two independent Eulerian and Lagrangian grid systems. This approach simplifies the modeling of curved boundaries of complex shapes. Many studies have been conducted to improve IBM accuracy, including the restoring force \cite{beyer1992analysis}, feedback forcing \cite{goldstein1993modeling}, and multidirect forcing methods \cite{breugem2012second}.

Although IBM was originally specifically designed for the Navier-Stokes level, IBM coupled with LBM has received increasingly more attention in recent years \cite{kang2011direct, thorimbert2018lattice, cheylan2021immersed}. The first combination of IBM and LBM was proposed by Feng and Michaelides \cite{feng2004immersed} in 2004. Kang and Hassan then proposed and estimated interface schemes for the direct-forcing immersed boundary method to address the problem of stationary boundaries \cite{kang2011direct, kang2011comparative}. For moving boundary problems, they suggested that the diffuse interface scheme is more suitable than the sharp interface scheme, as the latter can produce spurious oscillations. In the diffuse interface scheme, the forcing points are smoothly evolved along the boundaries of the immersed object. As a result, the diffuse interface scheme is appreciated more than in the intermediate Reynolds number ($Re$) range between $O(10^1)$ and $O(10^4)$, where most flying insects and birds appear \cite{shyy2016aerodynamics}. 

Previous research often focuses on accurate improvement of the numerical model in the uniform Cartesian mesh, while efficient management of the data structures for parallelization is often ignored \cite{tian2011efficient, feng2004immersed, zhu2021numerical}. A regular mesh system requires a large amount of memory for data storage and an enormous computational cost in time for the computation of each iteration. In addition, to ensure grid resolutions higher than those of adjacent regions, special attention is needed for regions near wall surfaces and areas of interest in the vortex wake \cite{moin1998direct, cao2022topological}. This condition ensures that the ratio of the local grid size to the Kolmogorov length scale is sufficiently small to capture the boundary layer and high-frequency eddies. The construction of a brilliant mesh generation framework is necessary for simulations that require high-performance computing. Previously, Nakahashi \cite{nakahashi2004building} introduced the building cube method for high-resolution flow calculations. The robustness and effectiveness of this method have been confirmed by subsequent studies \cite{nakahashi2006three, ishida2008efficient}. Recently, Duong et al. \cite{duong2022low} proposed a topology-confined block refinement for the two-dimensional LBM problem. This method can be easily extended for three-dimensional problems and parallel computation. 

In this paper, we propose a robust and efficient incompressible fluid solver based on the combination of LBM, IBM, and topology-confined block refinement framework for unbounded and bounded flows around stationary and flexible boundaries. The remaining part of the article is organized as follows: Section \ref{m} expresses the numerical methods based on the lattice Boltzmann method and the immersed boundary method combined with topology-confined block refinement used in this study. Section \ref{v} examines the accuracy and efficiency of the proposed framework using a variety of case studies from immersion objects to flexible wing motions. Finally, the main conclusions are summarized in Sect. \ref{c}.

\section{Methodology} \label{m}
This paper focuses on the development of an incompressible fluid solver. We do this by combining the lattice Boltzmann method (LBM), immersed boundary method (IBM), and topology-confined block refinement. The lattice Boltzmann model for incompressible flow is presented in Sect. \ref{m.lbm}. The method known as the multidirect forcing-immersed boundary method is subsequently introduced in Sect. \ref{m.ibm}. The article describes a framework that utilizes parallel topology-confined block refinement and nonuniform-grid interface interaction. The details of this framework may be found in Sect. \ref{m.topology} and Sect. \ref{m.nonuniform_grids}. Section \ref{m.implementation} provides a detailed description of the computational implementation.

\subsection{Lattice Boltzmann model for incompressible flow}\label{m.lbm}
\begin{figure}
    \centering
    \includegraphics[width=0.4\textwidth]{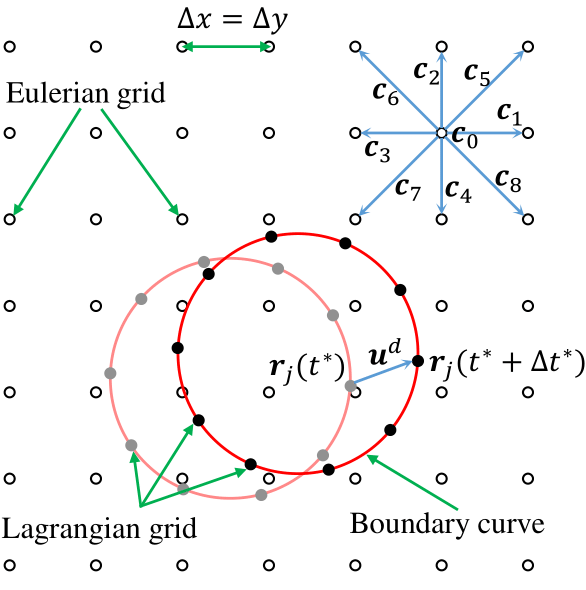}
    \caption{Sketch of 2D lattice space based on Eulerian grid and related illustrations of boundaries based on Lagranian grid. The circles represent particles in LBM. The colors matte red and solid red represent the boundary curves at time $t^*$ and $t^*+\Delta t^*$, respectively.
}
    \label{fig.1}
\end{figure}

The Boltzmann lattice model has been successfully utilized in a wide range of applications, including both two-dimensional (2D)\cite{duong2024near} and three-dimensional (3D)\cite{hafen2023numerical} scenarios. In this work, we focus on using the 2D model to illustrate the properties of the numerical model. In this 2D lattice Boltzmann model, the particle spacing ($\Delta x$) and the corresponding time step ($\Delta t$) are both set to 1. At each particle location $\boldsymbol{x}$, a discrete set of velocities $\boldsymbol{c}_i$ is defined. The two-dimensional discretization of these velocities is written as follows:
\begin{equation}
    \begin{aligned}
        c_{xi}=(0, 1, 0,-1, 0, 1,-1,-1, 1),\\
        c_{yi}=(0, 0, 1, 0,-1, 1, 1,-1,-1),
    \end{aligned}
\end{equation}
where the subscript $i$ stands for one of the nine directions linking neighbor particles, as shown in Fig. \ref{fig.1}. Thus, these form the discrete velocity set called D2Q9. In 3D space, two discrete velocity sets D3Q19 and D3Q27 are commonly used. In the current study, the Lattice Boltzmann equation for incompressible flow is expressed as follows \cite{bhatnagar1954model,kruger2017lattice}:
\begin{equation}\label{m.lbm.LB-equation}
    \Lambda_i(\boldsymbol{x}+\boldsymbol{c}_i\Delta t,t+\Delta t)=\Lambda_i (\boldsymbol{x},t)-\frac{1}{\tau}\left(\Lambda_i (\boldsymbol{x},t)-\Lambda_i^{\mathrm{e}} (\boldsymbol{x},t)\right) + F_i(\boldsymbol{x},t),
\end{equation}
where $\Lambda$ is the distribution function (DF) of the population related to the macroscopic space, $\Lambda^{e}$ is the equilibrium population and $\tau$ is the relaxation time. $F$ denotes the forcing term obtained from the extra force ($\boldsymbol{F}$), such as gravity or body force. According to Gou \cite{guo2002lattice}, the forcing term can be determined as follows.
\begin{equation}\label{forcing_term}
    F_i(\boldsymbol{x})=w_i \left(1-\frac{1}{2\tau}\right)\left[\frac{\boldsymbol{c}_i\cdot\boldsymbol{F}(\boldsymbol{x})}{c_s^2} +\frac{(\boldsymbol{c}_i\cdot\boldsymbol{u}(\boldsymbol{x}))(\boldsymbol{c}_i\cdot\boldsymbol{F}(\boldsymbol{x}))}{c_s^4}-\frac{\boldsymbol{u}(\boldsymbol{x})\cdot\boldsymbol{F}(\boldsymbol{x})}{c_s^2}\right].
\end{equation}
which satisfies the following requirements.
\begin{equation}
    \sum_i F_i(\boldsymbol{x}, t)=0,
\end{equation}
\begin{equation}
    \sum_i \boldsymbol{c}_i F_i(\boldsymbol{x}, t)=\left(1-\frac{1}{2 \tau}\right) \boldsymbol{F}(\boldsymbol{x}, t) .
\end{equation}

The computation process of Eq. \ref{m.lbm.LB-equation} is divided into two parts: collision and streaming. In the collision operation, the right-hand side of Eq. \ref{m.lbm.LB-equation} is implemented at the cell centers of a uniform Cartesian grid, where the particle population is located. The discrete post-collision DFs ($\Lambda_i^\star$) is given by
\begin{equation}\label{m.lbm.collision-equation}
    \Lambda_i^\star (\boldsymbol{x},t)=\Lambda_i (\boldsymbol{x},t)-\frac{1}{\tau}\left(\Lambda_i (\boldsymbol{x},t)-\Lambda_i^{\mathrm{e}} (\boldsymbol{x},t)\right) + F_i(\boldsymbol{x},t).
\end{equation}
For the LBM, the discrete equilibrium DFs can be expressed as
\begin{equation}\label{feq}
    \Lambda_i^{\mathrm{e}}(\boldsymbol{x})=w_i \rho(\boldsymbol{x})\left[1+\frac{\boldsymbol{c}_i\cdot\boldsymbol{u}(\boldsymbol{x})}{c_s^2} +\frac{(\boldsymbol{c}_i\cdot\boldsymbol{u}(\boldsymbol{x}))^2}{2c_s^4}-\frac{\boldsymbol{u}^2(\boldsymbol{x})}{2c_s^2}\right],
\end{equation}
where $w_{0}=4/9$, $w_{1-4}=1/9$, and $w_{5-8}=1/36$ correspond to weighting factors of linking direction, $c_s=1/\sqrt{3}$ indicates the speed of sound in lattice space, $\rho$ and $\boldsymbol{u}$ are local macroscopic fluid density and velocity, respectively. These macroscopic variables are computed using weighted sums known as moments of $\Lambda_i$ \cite{shu2014development}:
\begin{equation}\label{correct_u}
    \rho(\boldsymbol{x})=\sum_i \Lambda_i(\boldsymbol{x}), \quad \boldsymbol{u}(\boldsymbol{x})=\frac{1}{\rho(\boldsymbol{x})}\sum_i \boldsymbol{c}_i \Lambda_i(\boldsymbol{x})+\frac{\boldsymbol{F}(\boldsymbol{x})}{2\rho(\boldsymbol{x})}.
\end{equation}
Intrinsic average pressure is related to the speed of sound, given by $p(\boldsymbol{x})=\rho(\boldsymbol{x}) c_s^2$. The relaxation time is given through the fluid kinematic viscosity $\nu$ by the following equation:
\begin{equation}
    \nu = c_s^2 \left(\tau-\frac{1}{2}\right).
\end{equation}
The Reynolds number is defined by $Re=U_0 L/\nu$ with $L$ representing the characteristic length and $U_0$ representing the reference velocity of the fluid field. To determine the incompressible flow condition, $U_0$ is commonly set to ensure condition $U_0/c_s \leq0.1$ at the beginning of simulations. Normally, we recommend a value of $U_0=0.1$ or 0.05 for specific situations where higher stability is needed. During the collision process, multiple-relaxation time (MRT)\cite{du2006multi} and large-eddy simulation (LES)\cite{malaspinas2014wall} can be applied when the kinematic viscosity reduces for simulations of high Reynolds numbers.

After the collision, the DF population is updated by performing the streaming process. This process uses the post-collision DF populations of neighboring particles to determine the new states of the DF. The computation process is governed by the following formulas: 
\begin{equation}\label{streaming} 
    \Lambda_i(\boldsymbol{x}+\boldsymbol{c}_i\Delta t,t+\Delta t)=\Lambda_i^\star(\boldsymbol{x},t). 
\end{equation}
The streaming process is implemented in the lattice link between two adjacent fluid particles, where the discrete velocity components are located. However, for the end boundaries of the computational domain, the streaming process is not fully performed. To overcome this, interpolated schemes are used to generate the interpolated bounce-back method for the wall boundary condition and inflow, outflow, symmetry/free-slip, non-slip, and periodic for domain boundary conditions,  similar to conventional numerical methods. Interested readers are recommended to refer to our previous work\cite{duong2022low} for more details.

\subsection{Immersed boundary method}\label{m.ibm}
The immersed boundary method (IBM) is a scheme established by using both Eulerian and Lagrangian mesh systems. In the LBM, a particle population, which stores the values of the DFs and quantities, is used to solve the Lattice-Boltzmann equation, as indicated in Eq. \ref{m.lbm.LB-equation}. This population is regularly distributed at the cell center of a uniform Eulerian grid system. For the IBM, the rigid geometry surface is discretized and represented by marker points $\boldsymbol{r}_j$ in the Larangarian system, as shown in Fig. \ref{fig.1}. In this study, the distance between two adjacent marker points is set to ensure that they are equal to the spacing of the grid, $|\boldsymbol{r}_{jk}|=|\boldsymbol{r}_{j}-\boldsymbol{r}_{k}|=\Delta x$. 

To describe motions from spatio-temporally resolved numerical computation, the movement of boundaries is discretized in time. The position of a marker point is assumed to be $\boldsymbol{r}_j(t^*)$ at time $t^*$ with the superscript $^*$ representing the nondimentional time. After a time step $\Delta t^*$, the marker point is placed at $\boldsymbol{r}_j(t^*+\Delta t^*)$. When the rigid body moves in time steps, Lagrangian points translocate relatively with the desired velocity $u_j^\mathrm{d}=(\boldsymbol{r}_j(t^*+\Delta t^*)-\boldsymbol{r}_j(t^*))/\Delta t$. To link between two systems, the force density term is used to mimic the moving boundary condition. In this study, the force density around the marker points is obtained from the boundary force term, which is calculated by
\begin{equation}\label{f_lagrange}
\boldsymbol{F}_j=2\rho \frac{\boldsymbol{u}^\mathrm{d}_j-\boldsymbol{u}_j}{\Delta t}
\end{equation}
where $\boldsymbol{u}_j$ is the interpolated velocity obtained by interpolating from the uncorrected velocity of fluid particles around the Lagrangian point $\boldsymbol{r}_j$. This interpolation is governed by the following equation:
\begin{equation}\label{interpolate_Uj}
    \boldsymbol{u}_j=\sum \boldsymbol{u}_{\boldsymbol{x}} \mathcal{D} (\boldsymbol{x}-\boldsymbol{r}_j)\Delta x^2,
\end{equation}
with $\mathcal{D}$ denoting discretised version of the Dirac delta function given in 2D-space by
\begin{equation}
    \mathcal{D}(\boldsymbol{x})=\phi(x)\phi(y)/\Delta x^2,
\end{equation}
where $\phi$ is the kernel function. In this study, a stencil of $\phi$ is chosen to satisfy Peskin's claims:
\begin{equation}
\phi(x)= \begin{cases}\frac{1}{8}\left(3-2|x|+\sqrt{1+4|x|-4 x^2}\right) & 0 \leq|x| \leq \Delta x, \\ \frac{1}{8}\left(5-2|x|-\sqrt{-7+12|x|-4 x^2}\right) & \Delta x \leq|x| \leq 2 \Delta x, \\ 0 & 2 \Delta x \leq|x|.\end{cases}
\end{equation}
After computing the boundary force, the force density in particles around a point $\boldsymbol{r}_j$ is calculated by the equilibrium spreading distribution governed by the following equation:
\begin{equation}\label{spread_Fj}
    \boldsymbol{F}_{\boldsymbol{x}} =\sum \boldsymbol{F}_j \mathcal{D} (\boldsymbol{x}-\boldsymbol{r}_j)\Delta s_b.
\end{equation}
The total force acting on the rigid body's surface can be obtained by
\begin{equation}
    \boldsymbol{F}_{\mathrm{s}}=-\sum_j \boldsymbol{F}_j \Delta s_b=-\sum_{\boldsymbol{x}} \boldsymbol{F}_{\boldsymbol{x}} \Delta x^2,
\end{equation}
where $\Delta s_b$ indicates the arc length of the boundary segment.

\subsection{Topology-confined block refinement}\label{m.topology}
\begin{figure}
    \centering
    \includegraphics[width=0.55\linewidth]{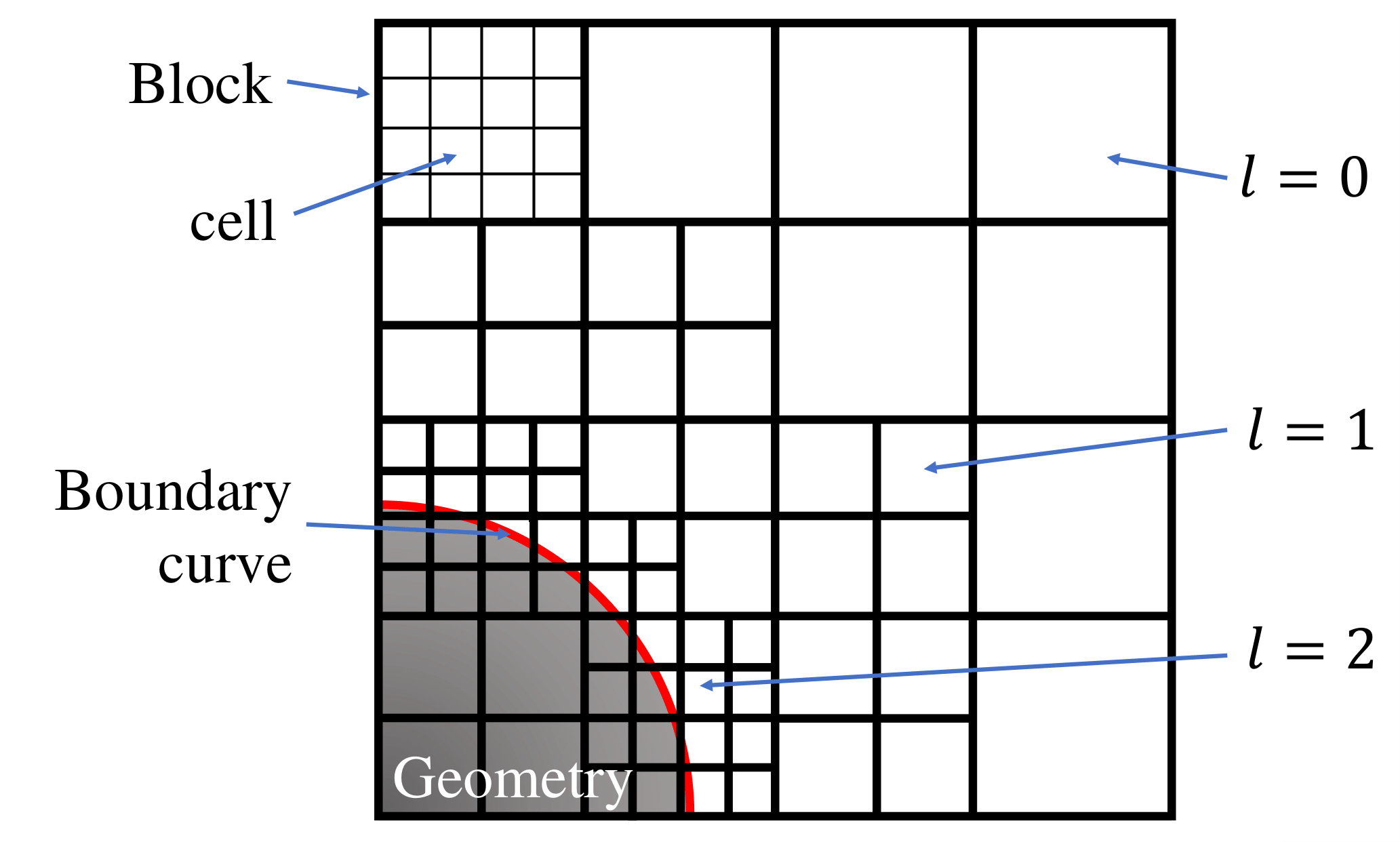}
    \caption{Illustration of the topology-confined block structure with three-level refinement.}
    \label{fig.2}
\end{figure}

In order to minimize the amount of computing resources required, we have created a system that utilizes parallel topology-confined block refinement. This system incorporates a distributed data structure that is especially optimized for parallel computations. In this case, the computational domain is partitioned into smaller parts that need varying grid resolutions. Each region is divided into regular square-shaped computing zones using square blocks, as seen in Fig. \ref{fig.2}. The grid spacing in various regions is dictated by the presence of flow-constrained areas, such as regions next to walls or specific flow regions that are necessary as per user requirements. The spatial variation is denoted by a refinement indicator $l$ depending on levels. Hence, the variable $l$ ranges from 0 to $l=m-1$, with $m$ being the required maximum degree of mesh refinement. The block information required for connecting to neighboring blocks is kept in the cell center of the uniform Cartesian grid used in each block. This information includes the block's index and rank, coordinates, spatial size, cell number, grid refinement level, and neighboring block connection information. These offer extensive guidelines within the field of parallel computing.

% To reduce computational costs, we developed a framework of parallel topology-confined block refinement with a distributed data structure specifically designed for parallel computations. Here, the computational domain is divided into smaller regions requiring different levels of the grid. In each region, square blocks are used to divide these domains into regular square-shaped computational zones, as shown in Fig. \ref{fig.2}. Grid spacing in different regions is determined by the flow-restricted region, such as regions near walls or specific flow regions required by users. This spatial difference is represented by a level-based refinement indicator $l$. Consequently, the value of $l$ ranges from 0 to $l=m-1$, where $m$ is the number of desired maximum levels of mesh refinement. To connect to neighboring blocks, block information, including block's index and rank, coordinates, spatial size, cell number, grid refinement level, and neighboring block linking information, is stored at the cell center of the uniform Cartesian grid, which is adopted in each block. These provide comprehensive rules within the parallel computing environment.

The numerical studies conducted by Kamatsuchi\cite{kamatsuchi2007turbulent} and Ishida et al.\cite{ishida2008efficient} indicate the effectiveness and robustness of this management method in handling the transition of information between refined and unrefined blocks. To achieve parallelism, the distribution of blocks at various refinement levels is carried out using a load-balanced linear distribution method that relies on the principles of space-filling curves\cite{bader2012space}. Block populations are linked to nodes or processors and subsequently sent to them for management. This method is executed using the MPI (Message Passing Interface) environment, which is specifically built for parallel computing systems. Once the block data has been sent to each node, the OpenMP (Open Multi-Processing) thread parallelization is used to perform separate tasks in a numerical manner.

% The efficiency and robustness of this management strategy on informational transition processes between refined and unrefined blocks is demonstrated in the numerical studies by Kamatsuchi\cite{kamatsuchi2007turbulent} and Ishida et al.\cite{ishida2008efficient}. For parallelism, the block distribution at different refinement levels is performed using a load-balanced linear distribution algorithm based on the theory of space-filling curves\cite{bader2012space}. Block populations are associated with nodes/processors and then sent to them to manage. This procedure is implemented with the MPI (Message Passing Interface) environment, which is specially designed to work on parallel computing architectures. After distributing block data to each node, independent workloads are numerically performed by OpenMP (Open Multi-Processing) thread parallelization. 
\subsection{Nonuniform-grid interface interaction}\label{m.nonuniform_grids}
Topology-confined block refining minimizes processing resources but results in non-uniform grids due to the difference in size between blocks of various grid levels. Implementing the LBM might result in disputes at the interfaces between blocks. In order to address this issue, a method is employed to handle the interaction of the DFs at the boundaries, allowing for the implementation of LBM on grids that are not uniform. Therefore, the block structure produced by topology-confined block refinement guarantees that the largest difference in grid levels between any two adjacent blocks is exactly 1. In addition, every block possesses two halo (ghost) layers that span over all faces, edges, and vertices.

% Although the use of topology-confined block refinement reduces computational resources, it creates a difference in size between blocks of different grid levels, known as non-uniform grids. This can cause conflicts at the interfaces between blocks when implementing the Lattice Boltzmann Method (LBM). To solve this problem, a strategy for the interaction of the distribution function at the interfaces is adopted to implement LBM on non-uniform grids. As a result, the block structure generated from topology-confined block refinement ensures that the maximum level difference between any two blocks of different grid levels is equal to 1. Additionally, each block has two halo (ghost) layers that extend across all faces, edges, and vertices.

\begin{figure}
    \centering
    \includegraphics[width=1.0\linewidth]{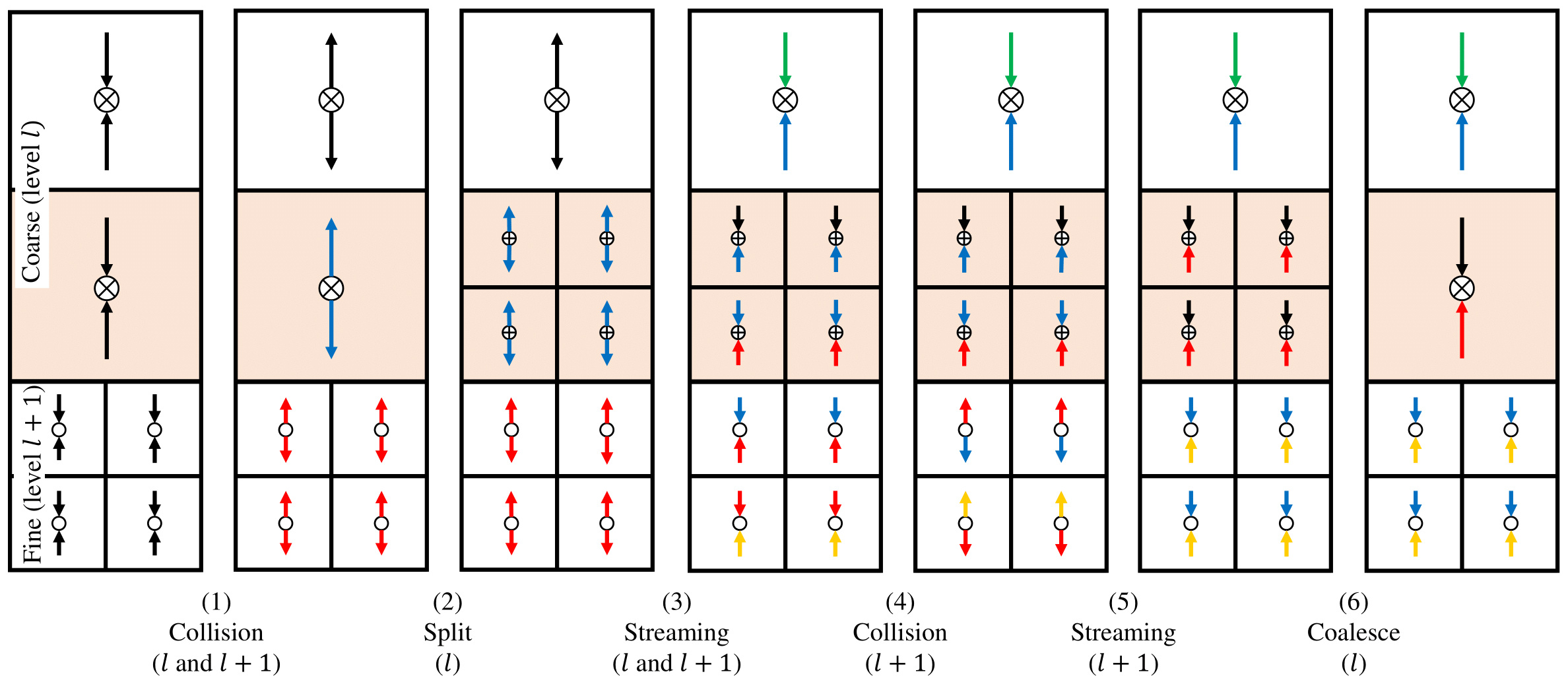}
    \caption{Illustration of nonuniform-grid interface interaction.}
    \label{fig.3}
\end{figure}

Figure \ref{fig.3} illustrates the procedure that implements LBM at the interfaces between two blocks with different grid levels. The technique was first proposed by Rohde et al. \cite{rohde2006generic} and later implemented for simulations on high-performance supercomputers in the work done by Schornbaum\cite{schornbaum2018block}. The present study used a modified version of Schornbaum's approach, using a smaller quantity of halo cells. Now, we examine a pair of levels, one called ``coarse" ($l$) and the other called ``fine" ($l+1$). At first, the simulation is in an initial or temporary state, when the time step is completely finished. The first step of the technique involves executing the collision process at all levels. Next, the border cell layer in the coarse block is relocated and divided into two halo layers in the fine block, as shown by the arrow in step (2). The streaming operation is thereafter performed in both blocks, including the two halo levels of the fine grid (3). During stages (4) and (5), an additional collision and streaming process takes place, specifically limited to the fine grid. The streaming procedure once again involves the utilization of the two halo layers. Ultimately, the values of the post-collision DFs in these two halo layers are merged and sent to the coarse grid in step (6).
% Figure \ref{fig.3} illustrates the algorithm implementing LBM at interfaces among two blocks of different grid levels. The approach was initially suggested by Rohde et al. \cite{rohde2006generic} and subsequently applied for simulations on extreme-scale supercomputers in the study conducted by Schornbaum\cite{schornbaum2018block}. In this study, we employ the modified version of Schornbaum's approach, but with a reduced number of halo cells. Now, we consider two blocks of levels, coarse ($l$) and fine ($l+1$). Initially, the simulation is in an initial or temporal state, where the time step is fully completed. Step (1) of the algorithm begins by performing the collision process at all levels. Then, a border cell layer in the coarse block is transferred and split into two halo layers in the fine block, as indicated by the arrow in step (2). The streaming process is then carried out in both blocks, including the two halo layers of the fine grid (3). In steps (4) and (5), another collision and streaming step is performed, but this time only in the fine grid. The streaming process again includes the two halo layers. Finally, the values of the post-distribution function in these two halo layers are coalesced and transferred to the coarse grid in steps (6). To summarize the methodology, Figure \ref{fig.4} illustrates the entire computational framework proposed within the standard time step.

\subsection{Implementation}\label{m.implementation}
\begin{figure}
    \centering
    \includegraphics[width=0.8\linewidth]{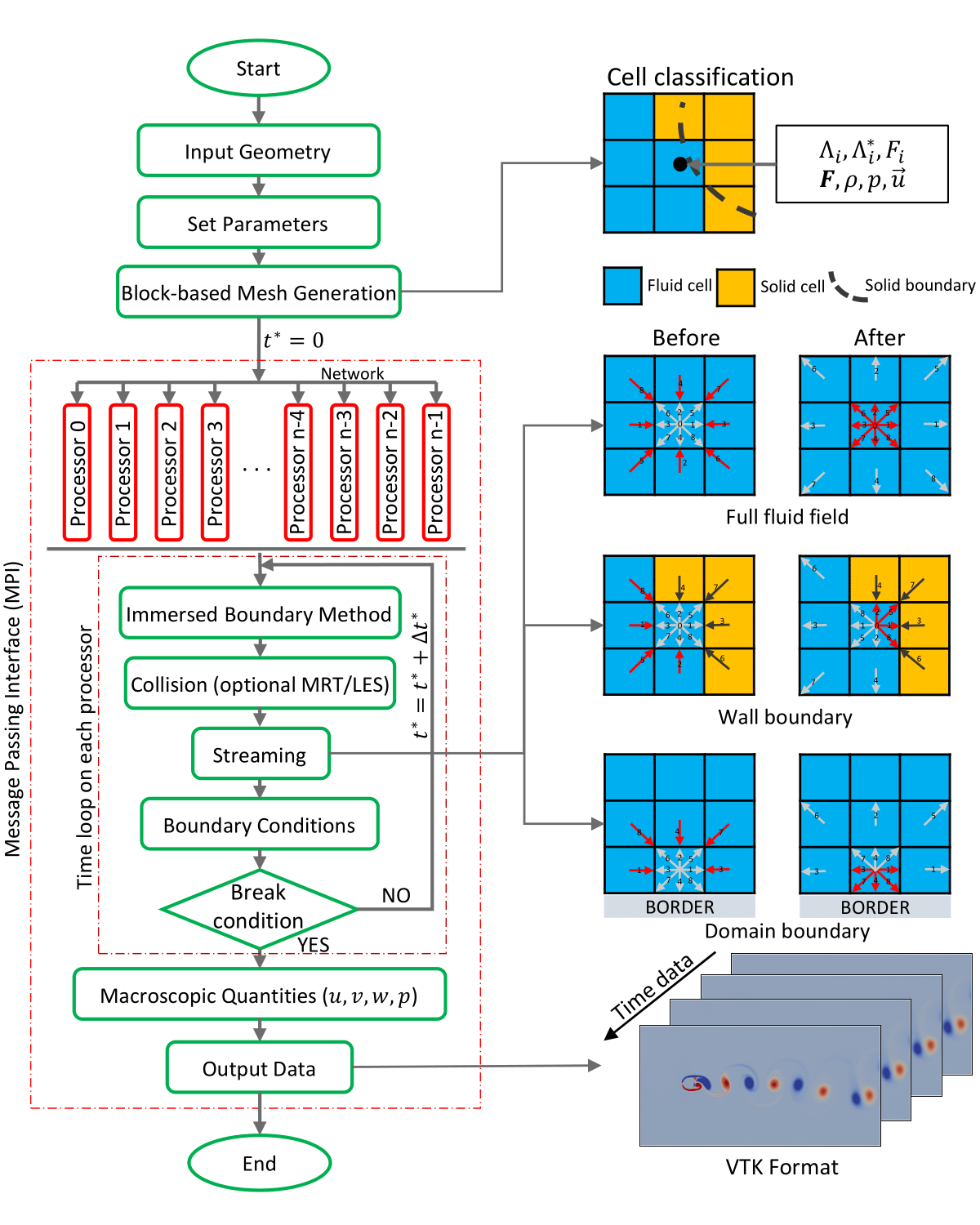}
    \caption{Flow chart of the computational framework (left); Illustration of cell classification, streaming processes, and VTK-format-based output files (right).}
    \label{fig.4}
\end{figure}

To provide a comprehensive understanding of the computational implementation, Figure \ref{fig.4} visually depicts the entirety of the computational framework. Initially, the solver is provided with the geometry data to establish the bounds for the model. Subsequently, users enter parameters related to motion modes and fluid flow qualities. Once all the necessary parameters are obtained, a mesh is created based on blocks using a framework that limits refinement to certain areas of the mesh. For static situations involving solid objects that do not involve the use of IBM, the cell population is categorized into fluid cells and solid cells. Afterwards, the data is initialized and sent to nodes/processors for computation within an MPI environment. The OpenMP thread parallelization technique is employed to execute LBM, IBM, and other operations inside the framework on each node or CPU. The IBM (Immersed Boundary Method) is initially employed here to update and supply quantities for the subsequent collision procedure using moving boundaries. It should be noted that the collision step can be carried out with the optional use of MRT and LES. The streaming process is incomplete when it comes to the wall and outer limits, resulting in the establishment of two boundary conditions. Once the program is finished or when there is a need to analyze macroscopic quantities, the data are exported and accessed using the VTK format\cite{vtkBook} for visualizations.

Algorithm \ref{alg.1} demonstrates the fundamental framework of the synchronous recursive procedure that depicts the program flow associated with LBM on nonuniform grids in the primary time loop on each processor. The function comprises two components corresponding to two stages of a meticulous level. For the purpose of enhancing legibility, functions that entail communication are denoted by the term ``Communication" in their respective function names. The remaining functions carry out subtasks of the LBM, such as the \textsc{CollisionStep} and \text{StreamingStep} functions. This algorithm serves as the algorithmic representation of the parallelization technique for the algorithms depicted in Figs. \ref{fig.3} and \ref{fig.4}. The algorithm is invoked with $l$ set to zero. Consequently, the simulation progresses by one coarse time step ($l$ equal to zero represents the coarsest level). The \textsc{UpdateVelocity} function is responsible for updating the velocity field and force term using the Immersed Boundary Method (IBM). These updated values are subsequently utilized in the \textsc{CollisionStep} function. The ``Communication" functions handle the transfer of data between grids at the same and other levels, including steps (2) and (6) shown in Fig. \ref{fig.3}. In the second streaming process, a function called \textsc{PreStreamingStep} is introduced to replicate the collision phase between steps (4) and (5) by reversing the DFs in the halo layers at the fine level. The \textsc{BoundaryCondition} function handles two types of boundary conditions that are generated by the streaming step.

\begin{algorithm}
\caption{Recursive algorithm illustrates the program flow for the LBM on nonuniform grids}\label{alg.1}
\begin{algorithmic}
\Function{SynchronousNonUniformTimeStep}{level $l$}
\State \textbf{call} \Call{UpdateVelocity}{$l$} \Comment{IBM-based updated velocity field}
\State \textbf{call} \Call{CollisionStep}{$l$}\Comment{collision step with force term}
\If{$l \neq m-1$}
    \State \textbf{recursive call} \Call{SynchronousNonUniformTimeStep}{$l+1$}\Comment{recursive call}
\EndIf
\If{$l \neq 0$}
     \State \textbf{call} \Call{CoarseToFineCommunication}{$l-1$, $l$}\Comment{coarse-to-fine DFs communication}
\EndIf
\State \textbf{call} \Call{SameLevelCommunication}{$l$}\Comment{same-level DFs communication}
\State \textbf{call} \Call{StreamingStep}{$l$}\Comment{streaming at interior part of block}
\State \textbf{call} \Call{BoundaryConditionStep}{$l$}\Comment{wall and outer boundary conditions}
\If{$l \neq m-1$}
     \State \textbf{call} \Call{FineToCoarseCommunication}{$l+1$, $l$}\Comment{fine-to-coarse DFs communication}
\EndIf
\If{$l = 0$}
     \State \Return{}\Comment{ending of the algorithm}
\EndIf
\State \textbf{call} \Call{UpdateVelocity}{$l$}\Comment{IBM-based updated velocity field}
\State \textbf{call} \Call{CollisionStep}{$l$}\Comment{collision step with force term}
\If{$l \neq m-1$}
    \State \textbf{recursive call} \Call{SynchronousNonUniformTimeStep}{$l+1$}\Comment{recursive call}
\EndIf
\State \textbf{call} \Call{PreStreamingStep}{$l$}\Comment{streaming at halo layers}
\State \textbf{call} \Call{StreamingStep}{$l$}\Comment{streaming at interior part of block}
\State \textbf{call}
\Call{BoundaryConditionStep}{$l$}\Comment{wall and outer boundary conditions}
\If{$l \neq m-1$}
     \State \textbf{call} \Call{FineToCoarseCommunication}{$l+1$, $l$}\Comment{fine-to-coarse DFs communication}
\EndIf
\State \Return{}\Comment{ending of the algorithm}
\EndFunction
\end{algorithmic}
\end{algorithm}

\section{Validation study} \label{v}
In this section, the current solver for the immersed boundary-lattice Boltzmann method is validated based on experimental and numerical data published by other authors. Specifically, five case studies are utilized to examine the solver for a wide variety of parameters, including shapes, Reynolds numbers, and motion types. We first verify second-order accuracy of the LBM model in Sect. \ref{v.1} and then examine the reliability of the hydrodynamic quantities computed for the flow over an obstacle-embedded lid-driven cavity in Sect. \ref{v.2}. The effects of boundary conditions and topology-confined block structures on fluid flow over an isolated circular cylinder are then investigated in Sect. \ref{v.2}. In Sect. \ref{v.4}, we use this framework to probe the fluid dynamic properties of the single pitch-up motion.  Finally, large-scale flapping wing motions of both amplitude and frequency are investigated in Sect. \ref{v.5} and Sect. \ref{v.6}, respectively.
\subsection{Taylor-Green Vortex Flow}\label{v.1}
\begin{figure}
    \centering
    \includegraphics[width=0.8\textwidth]{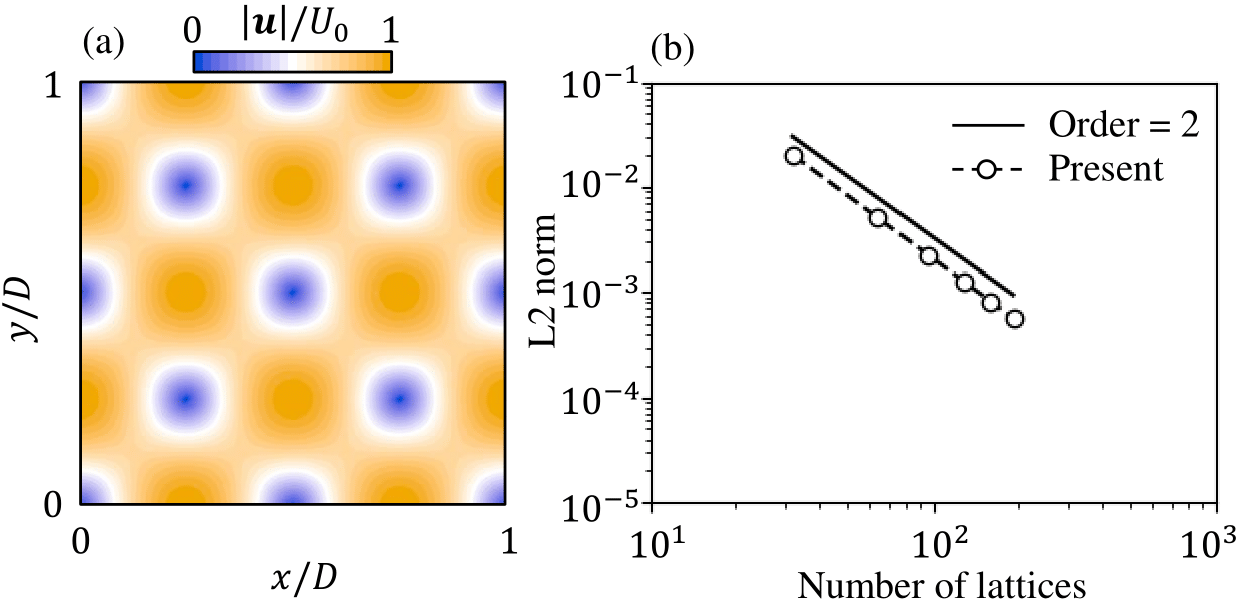}
    \caption{Initial velocity magnitude contours of 2D Taylor-Green vortex flow (a); Investigation of L2 norm of velocity component $u$ (b).}
    \label{fig.5}
\end{figure}

In this study, we first consider the decaying Taylor-Green \cite{taylor1937mechanism} vortex (TGV) flow to verify the accuracy of the present solver under the incompressible flow condition. TGV is a typical unsteady flow situation widely used as a benchmark problem to validate solvers with a fully periodic domain size $[0,1]$. Due to the symmetry property, we use two-dimensional flow investigations in this work. The evolution of the velocity and pressure components is expressed by an analytical solution, which is given by \cite{kruger2017lattice}
\begin{equation}
    \begin{aligned}
        u(\boldsymbol{x}, t)&=-U_0\cos (2\pi x) \sin (2\pi y) e^{-t/t_d}, \\
        v(\boldsymbol{x}, t)&=U_0\sin (2\pi x) \cos (2\pi y) e^{-t/t_d}, \\
        p(\boldsymbol{x}, t)&=p_0-\rho_0\frac{u_0^2}{4}(\cos (2\pi x)+\sin (2\pi y)) e^{-2 t /t_d}
    \end{aligned}
\end{equation}
with the vortex decay time $t_d=1/2\pi^2\nu$ defined by fluid viscosity $\nu$. The $u$ and $v$ denote the two components of velocity along directions $x$ and $y$ at location $\boldsymbol{x}$ and time $t$. The pressure average $p_0$ and the average density $\rho_0$ are set to 0 and 1, respectively. The initial flow state is defined by $\boldsymbol{u}(\boldsymbol{x}, 0)$ and $p(\boldsymbol{x}, 0)$, as shown in Fig. \ref{fig.5}(a); the initial density field is determined by $\rho(\boldsymbol{x}, 0)=\rho_0+p(\boldsymbol{x}, 0)/c_s^2$.

To study the accuracy, we performed all simulations at fluid viscosity $\nu=1/6$ in the periodic square domain $[0,1]\times[0,1]$. Five different resolutions ($N_x\times N_y$=$32\times 32$, $64\times 64$, $96\times 96$, $128\times 128$, $160\times 160$, and $192\times 192$) are used to examine spatial precision. The characteristic flow velocity ($U_0$) is set to ensure the same Reynolds numbers $Re=U_0 N_x/\nu=7.68$. The results obtained are compared with each other in one nondimensional time $t^*=tU_0/N_x=0.25$ to provide temporal accuracy. To determine the error norm $L_2$, the velocity component $u$ is used in this work due to the symmetry of the TGV. The velocity error is checked by comparing the present results with the analytical solution mentioned at the same time $t^*=0.25$. As shown in Fig. \ref{fig.5}(b), second-order accuracy is observed when increasing the number of lattices on the domain size. It is evident that the current SBoTFlow solver provides second-order accuracy for numerical computation under incompressible flow conditions, similar to other NSE solvers.
\subsection{Lid-driven cavity flow with embedded cylinder}\label{v.2}
\begin{figure}
    \centering
    \includegraphics[width=0.5\linewidth]{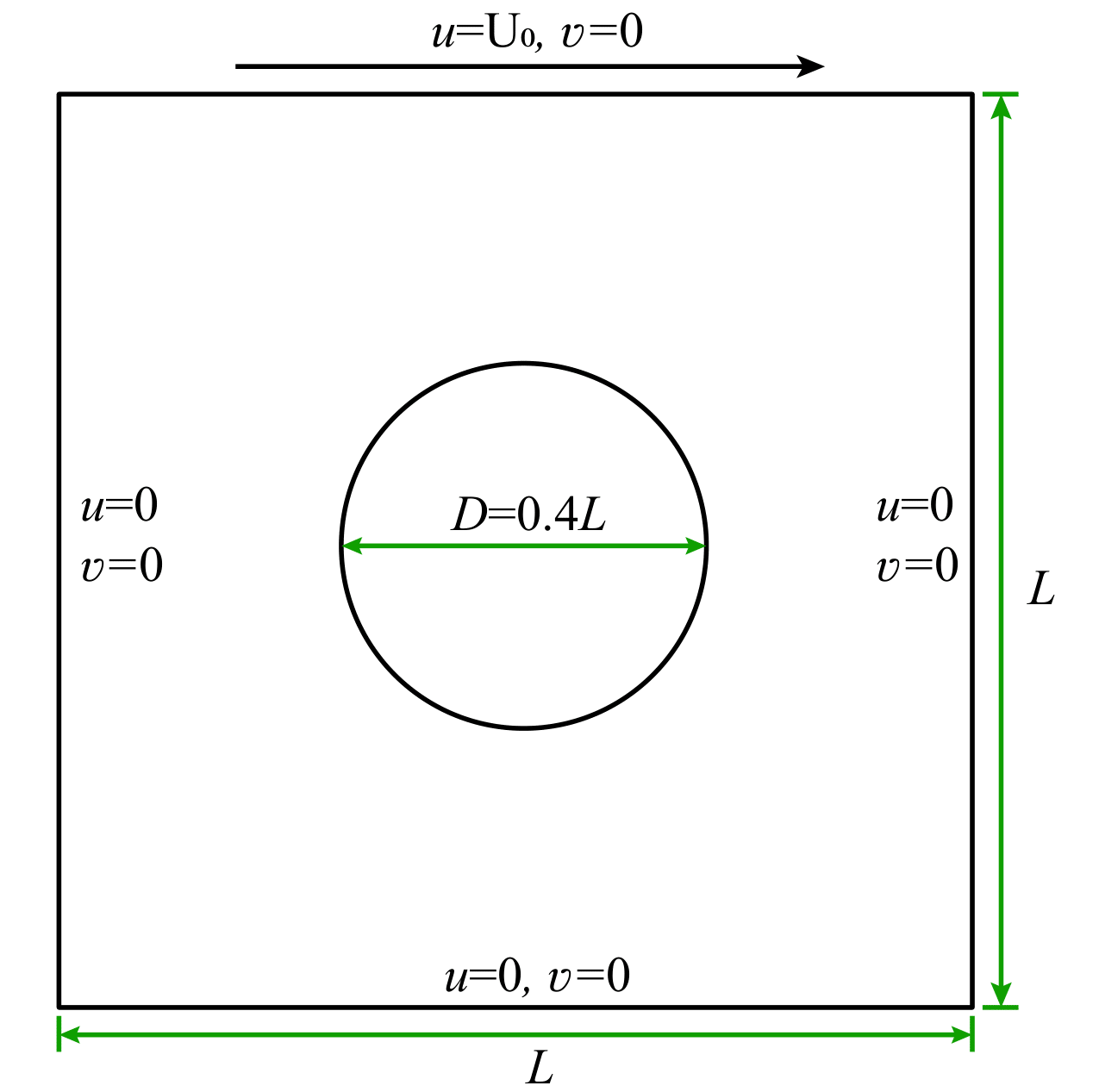}
    \caption{Sketch of the computational domain of Lid-driven cavity flow with a circular obstacle immersed at the center.}
    \label{fig.6}
\end{figure}

In this section, we consider the problem of a circular cylinder embedded at the center of the lid-driven cavity. The flow configuration and boundary conditions are the same as those of Cai et al. \cite{cai2017moving} and Rajan and Perumal \cite{rajan2021flow}. The cavity is determined by the square domain with side length $L=1$. The upper wall has a constant velocity $U_0=0.05$, which is considered the reference velocity, in the lattice unit. The bounce-back boundary condition for stationary walls is defined on the remaining walls, as shown in Fig. \ref{fig.6}. The diameter of the cylinder is $D=0.4L$. The cylinder's surface is set to non-slip condition by IBM. To compare with the results of Rajan and Perumal \cite{rajan2021flow}, a mesh of $300\times 300$ is used for three simulations with Reynolds numbers of $100$, $400$, and $1000$. In addition, we also used this setup for the interpolated bounce-back method to compare the immersed bounce method with the traditional boundary modeling method. Interested readers can find a detailed presentation of the interpolated bounce-back method in our other work \cite{duong2022low}. The convergence criterion of all simulations is defined as
\begin{equation}
    E=\frac{\sum_{\boldsymbol{x}} |\boldsymbol{u}(\boldsymbol{x},t+1000)-\boldsymbol{u}(\boldsymbol{x},t)|}{\sum_{\boldsymbol{x}} |\boldsymbol{u}(\boldsymbol{x},t+1000)|}\leq 10^{-6}.
\end{equation}

\begin{figure}
    \centering
    \includegraphics[width=1.0\linewidth]{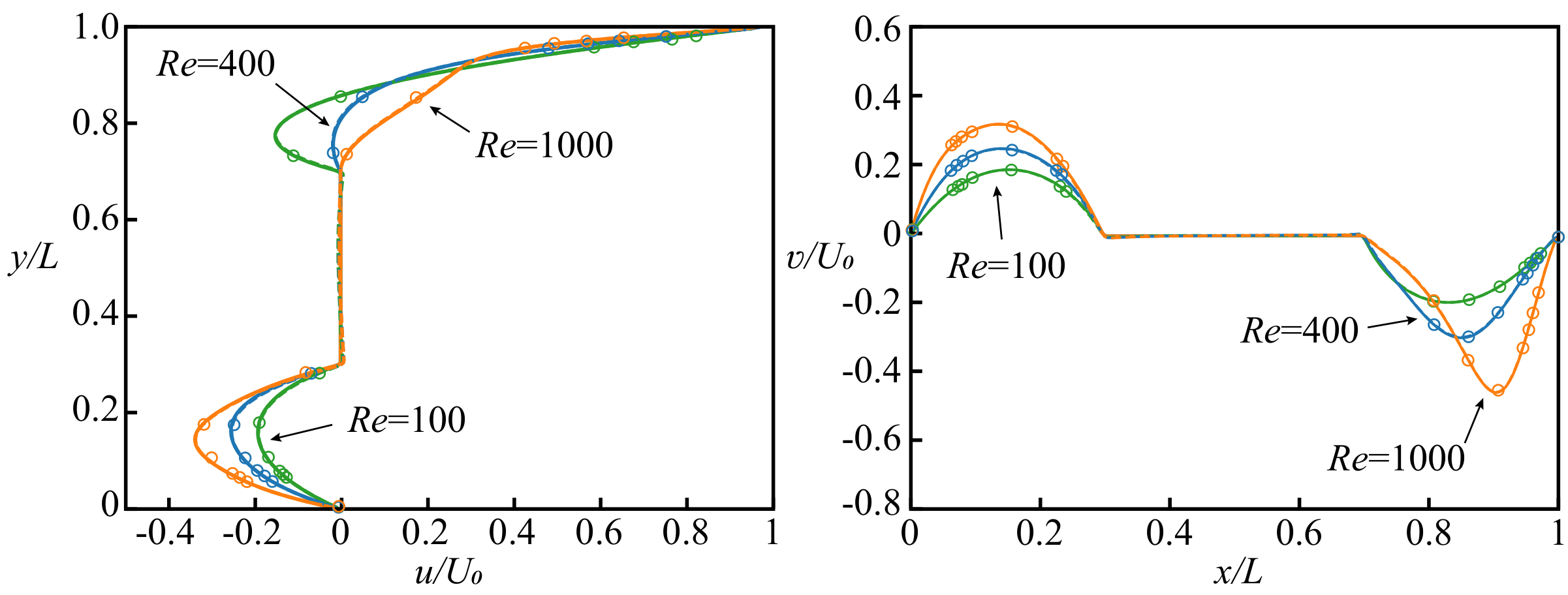}
    \caption{Velocity profiles across the cavity center. Solid line represents the SBoTFlow's results with the traditional interpolated bounce-back method; Dashed line represents IBM-based SBoTFlow's results; Open circle indicates numerical findings by Rajan and Perumal. \cite{rajan2021flow}}
    \label{fig.7}
\end{figure}

Figure \ref{fig.7} shows the values of the velocity components ($u$ and $v$) along the horizontal and vertical center lines of the cavity. The results obtained from both the interpolated bounce-back method and immersed boundary method are compared with the simulation data published in the work of Rajan and Perumal \cite{rajan2021flow}. Overall, the SBoTFlow-based results agree well with both the reference data and the results obtained from the interpolated bounce-back condition. A small deviation is observed in the boundary region due to the effect of the stencil function at the maximum Reynolds number of 1000. This inherent deviation of IBM is often ignored in almost all investigations because of tiny impacts on studies and their conclusions. 

\subsection{Flow past a stationary circular cylinder}\label{v.3}
\begin{figure}
    \centering
    \includegraphics[width=0.7\textwidth]{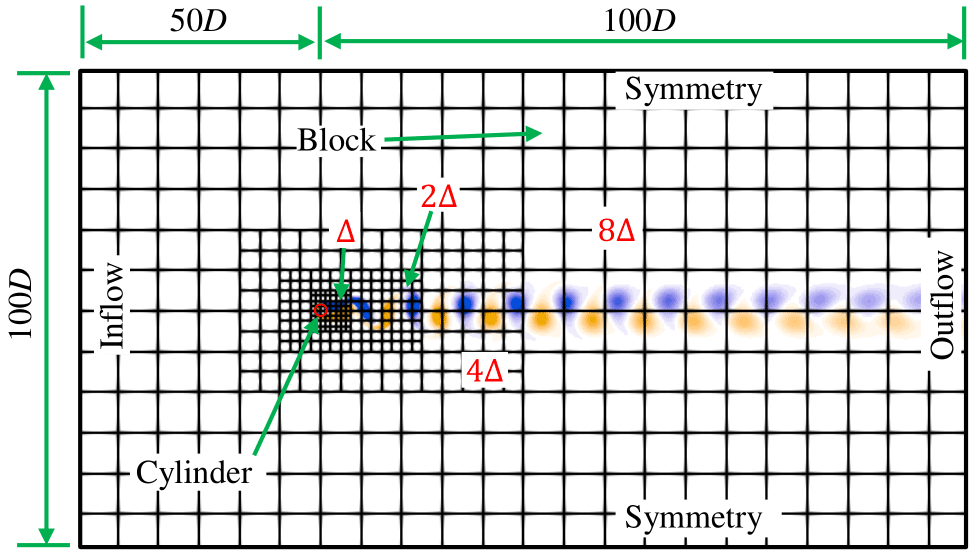}
    
    \caption{Schematic of the computational domain for flow past a stationary circular cylinder.}
    \label{fig.8}
\end{figure}

To determine the effects of the far boundaries and the topology-confined block refinement algorithm, we investigated the flow through a stationary circular cylinder with different Reynolds numbers of 100, 300, and 550. As shown in Fig. \ref{fig.8}, the size of the computational domain is $100D\times 150D$ with $D=1$ denoting the diameter of the cylinder. The open boundary condition is imposed at the outlet, whereas the equilibrium boundary condition is applied at the inlet. The symmetry boundary condition is set to two side boundaries. The surface of the cylinder is represented by a population of constantly distributed Lagrangian points with $\Delta s = \Delta$, with $\Delta=1/128$ representing the minimum grid spacing obtained from the finest grid region of the maximum level of grid refinement. In this work, four levels of grid refinement consisting of $\Delta$, $2\Delta$, $4\Delta$, and $8\Delta$ are established to satisfy the capture of wakes behind the cylinder. The free-stream velocity $U_0$ is set to $0.1$. In this case study, Reynolds numbers are determined based on the free-stream velocity and diameter of the cylinder, $Re=U_0 D/\nu$. As a result, the values of the kinetic viscosity $\nu$ depend on the Reynolds numbers.

In the case of $Re = 100$ and $Re =300$, the internal instability of the flow must be statistically evaluated. We focus on three parameters in particular: the average drag coefficient over time ($\overline{C}_{D}$), root mean square of the lift coefficient ($C_{L}^{\prime}$), and the Strouhal number ($St$), which are expressed by the following equations.
\begin{equation}\label{E21}
	\overline{C}_{D}=\frac{1}{N}\sum_{1}^{N}C_{D}, \quad C_{L}^{\prime}=\sqrt{\frac{1}{N}\sum_{1}^{N}\left(C_{L}-\overline{C}_{L}\right)^2}, \quad St=\frac{f \cdot D}{U_0}
\end{equation}
Here $C_D$ and $C_L$ indicate the instantaneous drag and lift coefficients. The overline symbol $-$ represents the average, while the symbol $^\prime$ denotes the root mean square of the hydrodynamic coefficients, with $N$ representing the number of measurements. Furthermore, $f$ represents the wake frequency calculated directly from the time history of the instantaneous lift coefficient. In Table \ref{tab.1}, the results of the hydrodynamic coefficients ($\overline{C}_{D}$, $C_{L}^{\prime}$, and $St$) are presented and compared with other literature. The values derived from the current calculation consistently fall within the range reported by previous studies.
\begin{table}

\caption{Comparison of hydrodynamic coefficients generated by flow past an isolated circular cylinder at $Re=100$ and $300$.}
\label{tab.1}
\centering
\begin{tabular}{lcccccc}
\hline\hline
\multicolumn{1}{c}{\multirow{2}{*}{Authors}} & \multicolumn{3}{c}{$Re=100$} & \multicolumn{3}{c}{$Re=300$} \\ \cline{2-7} 
\multicolumn{1}{c}{}  & $\overline{C}_{D}$    & $C_{L}^{\prime}$   & $St$   & $\overline{C}_{D}$   & $C_{L}^{\prime}$  & $St$   \\ \hline
Ghias et al. \cite{ghias2004non}  & 1.36     & 0.32     & 0.16   & 1.40     & 0.67     & 0.21   \\
De Palma et al.   \cite{de2006immersed} & 1.32     & 0.23     & 0.16   & -        & -        & -      \\
Rajani et al. \cite{rajani2009numerical} & 1.33     & 0.17     & 0.15   & 1.28     & 0.60     & 0.21   \\
Boukharfane et al.  \cite{boukharfane2018combined} & 1.36     & 0.25     & 0.16   & 1.26     & 0.62     & 0.21   \\
Vanna et al. \cite{de2020sharp}  & 1.32     & 0.22     & 0.16   & 1.34     & 0.63     & 0.21   \\
Present    & 1.33     & 0.24     & 0.16   & 1.31     & 0.62     & 0.21   \\ \hline\hline
\end{tabular}

\end{table}

\begin{figure}
    \centering
    \includegraphics[width=0.8\textwidth]{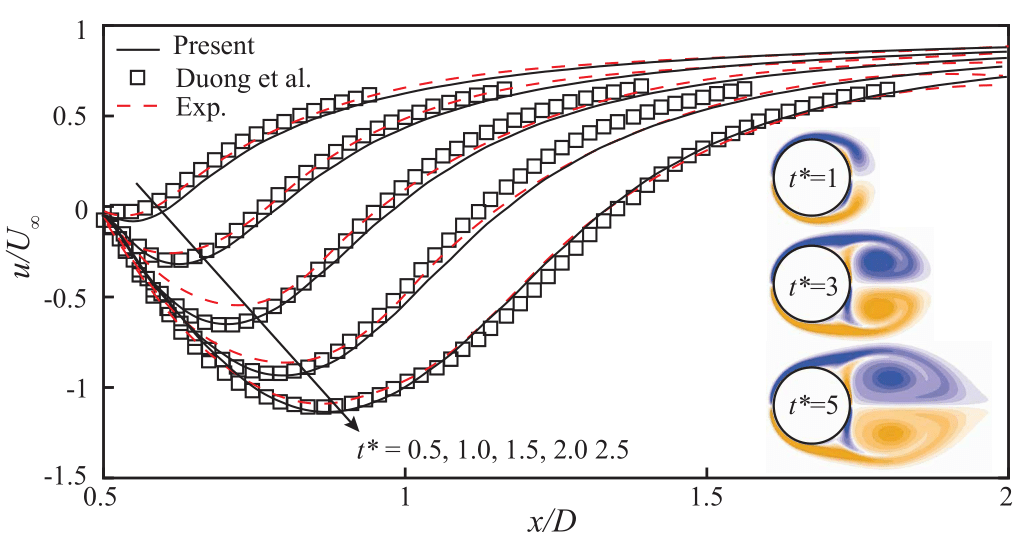}
    \caption{Horizontal centerline velocity profile and colored vorticity contours for flow through a circular cylinder with $Re=U_\infty D/\nu=550$ at different times. Symbols $\square$, red dashed line, and solid line denote experimental data \cite{bouard1980early}, the result obtained simulation\cite{duong2021fluid} and present result, respectively.}
    \label{fig.9}
\end{figure}

Figure \ref{fig.9} shows the distribution of the horizontal centerline velocity and the colored vorticity contours at different times $t^{*}=tU_0/D= 1.0, 3.0$, and $5.0$. The symbols $\square$ denote the experimental data under the same condition \cite{bouard1980early}. Red dashed lines and black solid lines indicate the simulation results obtained from the numerical calculation of the study by Duong et al.\cite{duong2021fluid} and the present work. Horizontal centerline velocities are compared in the recirculating flow region. The figure indicates that the results obtained from SBoTFlow agree well with the experimental data and the result obtained from the purely fast Lagrangian vortex method of Duong et al.\cite{duong2021fluid}. The evolution of the flow is quantitatively illustrated by a sequence of vorticity contours. Here, blue and yellow indicate clockwise and anticlockwise vorticities. The wake behind the circular cylinder develops symmetrically from the initial state to the later state of the simulation at $t^*=5$. A couple of vortex bubbles are formed and gradually evolve into larger bubbles during the flow evolution time. In the vorticity contour, the separation point is around $60^\circ$, which is consistent with the numerical and experimental results of Kim \cite{kim2012vortex} and Taneda \cite{taneda1956experimental}. These indicate that the present solver has successfully simulated static objects under incompressible flow conditions.
\subsection{Pitch-up maneuvered flat plate}\label{v.4}
\begin{figure}
    \centering
    \includegraphics[width=0.8\textwidth]{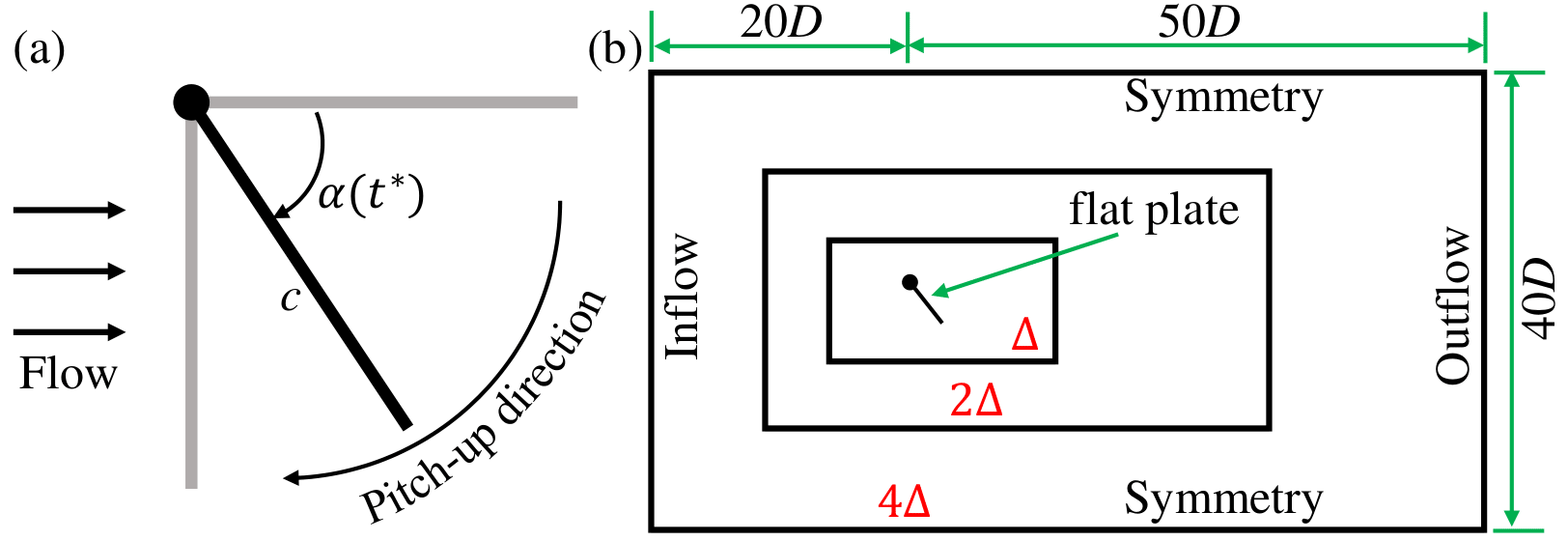}
    \caption{Sketch of plat-plate pitch-up movement (a); Schematic of the computational domain for flow over pitch-up maneuvered plat plate (b).}
    \label{fig.10}
\end{figure}

Understanding the fluid dynamics surrounding the motion of a rigid body is crucial in flow physics research. To address this issue, the present solver has been developed to simulate the movements of complex trajectories. In this study, we evaluate the solver by analyzing the flow behavior over a flat-plate geometry with square leading and trailing edges at a Reynolds number of 1000. The thickness of the flat plate is set to $2.3\%$, and the characteristic chord $c$ is 1, as depicted in Fig. \ref{fig.10}(a). Furthermore, the computational domain is defined as a rectangle with dimensions of $40D\times70D$. To achieve the smallest grid spacing $\Delta$, three refinement levels are employed. The boundary conditions are defined as symmetry, inflow, and outflow for two sides, inlet, and outlet, respectively. This approach is similar to the issue of fluid flow around a circular cylinder, as shown in Fig. \ref{fig.10}(b). The SBoTFlow solver is validated against the theoretical solution results of Ramesh et al. \cite{ramesh2014discrete} and the numerical results of Eldredge \cite{eldredge2007numerical} and Duong et al. \cite{duong2021fluid}.

The pitch ramp utilizes a smoothed ramp hold-return motion to achieve a pitch up from $0^\circ$ to $90^\circ$ around the leading edge. Eldredge et al.\cite{eldredge2009computational} and Wang and Eldredge\cite{wang2013low} have established a canonical formulation using a smoothing function defined as follows (Eq. \ref{eq.smoothing_function}): 
\begin{equation}\label{eq.smoothing_function} G(t^*)=\mathrm{log}\left[\frac{\mathrm{cosh}(a_{s} (t^*-t_1^*))}{\mathrm{cosh}(a_{s} (t^*-t_2^*))}\right]-a_s (t_1^* - t_2^*), \end{equation}
where $a_s$ is a free parameter that controls smoothing between kinematic intervals. The times $t_1^*$ and $t_2^*$ represent transitional moments during the pitch-up maneuver. $t_1^*$ marks the start of the pitch-up and the beginning of the deceleration, while $t_2^* = t_1^* +A/(2K)$ indicates the end of the pitch-up and the end of deceleration. Here, $A$ and $K$ are the pitching amplitude (in radians) and the nondimensional pitching rate, respectively. The time step is denoted as $t^*$. In this specific case study, $t_1^* = 1$ is chosen to ensure sufficient development of the boundary layers on the surface of the plate before rotation begins. It is important to note that $t_1^*$, $t_2^*$, and $t^*$ in Eq. \ref{eq.smoothing_function} have been normalized into nondimensional time parameters using $U_0/c$. The time history of the angle of attack is governed by the ramp-hold-return motion, expressed as follows:
\begin{equation}
    \alpha(t^*)=A\frac{G(t^*)}{\mathrm{max}(G(t^*))},
\end{equation}
where $\mathrm{max}(G(t^*))$ is approximately equal to $ 2a_s (t_2^* -t_1^*)$. In this work, the Reynolds number is examined as $Re=U_0 c/\nu=1000$ with free stream velocity $U_0=0.05$.  The parameters $A$, $a_s$, and $K$ are set to $\pi/2$, 11, and 0.2, respectively.
\begin{figure}
    \centering
    \includegraphics[width=1.0\textwidth]{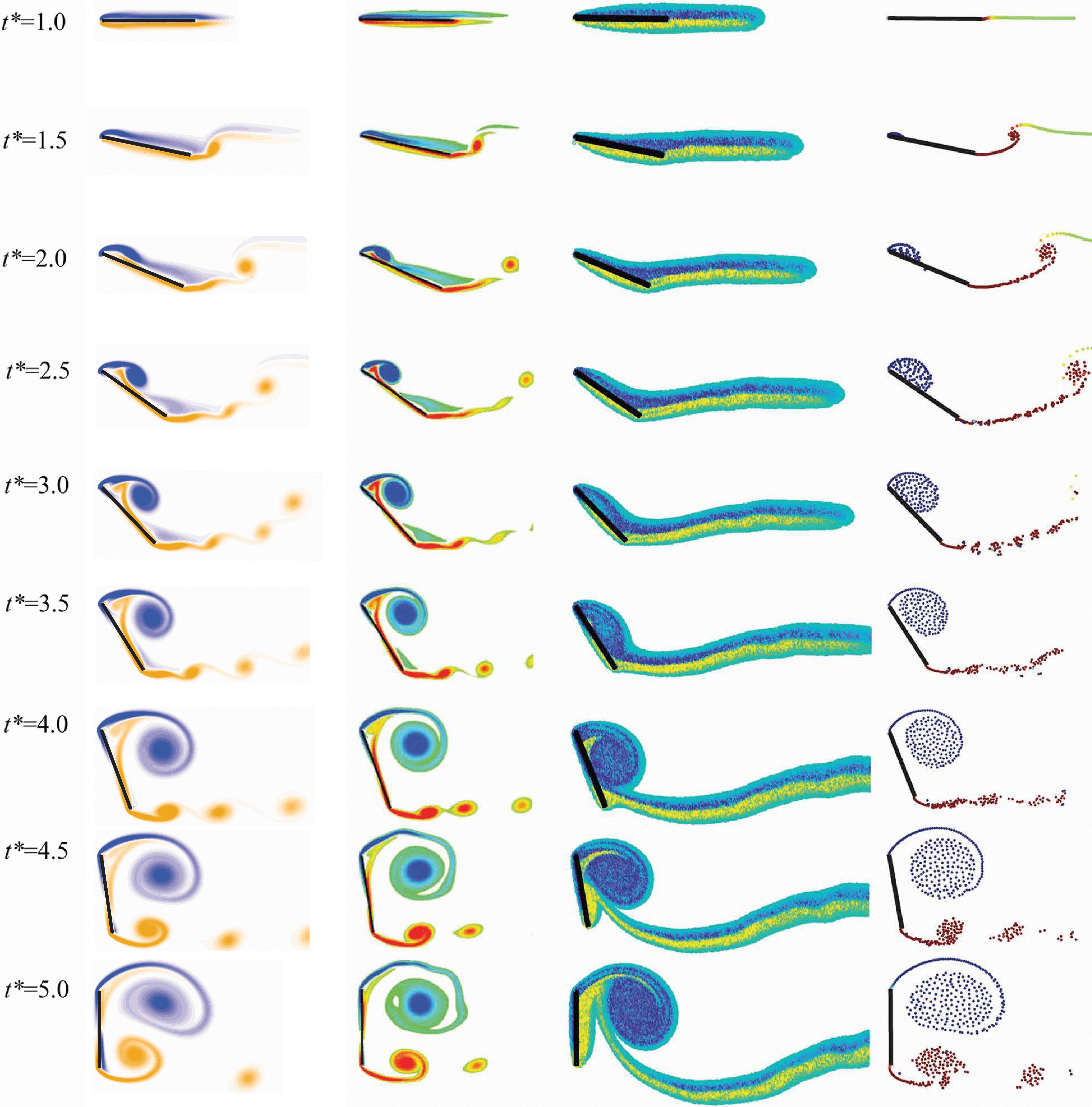}
    \caption{Comparion of vorticity contours between present results and previous computational and theory results. }
    \label{fig.11}
\end{figure}

Figure \ref{fig.11} presents an overview of the flow variation on the surface of a flat plate. The vorticity results obtained in this study are compared with those of previous research using numerical and theoretical methods. The nondimensional time evolution is considered from $t^*=1.0$ to $t^*=5.0$. The flow characteristics associated with the vorticity contours are shown for the current SBoTFlow (first column), Computational Fluid Dynamics (CFD) by Wang and Eldredge \cite{eldredge2007numerical} (second column), Fast Lagrangian Vortex Method (FLVM) by Duong et al. \cite{duong2021fluid} (third column), and LESP modulated Discrete Vortex Method (LDVM) by Ramesh et al. \cite{ramesh2014discrete} (fourth column). The formation and separation of the leading edge vortex (LEV) and the trailing edge vortex (TEV) are in excellent agreement with the results obtained from LDVM and CFD. The LEV is captured at the same position and magnitude as observed in the reference data during the pitching motion. The formation of hairpin structures at $Re=1000$ is clearly observed in the TEV.

\begin{figure}
    \centering
    \includegraphics[width=0.6\textwidth]{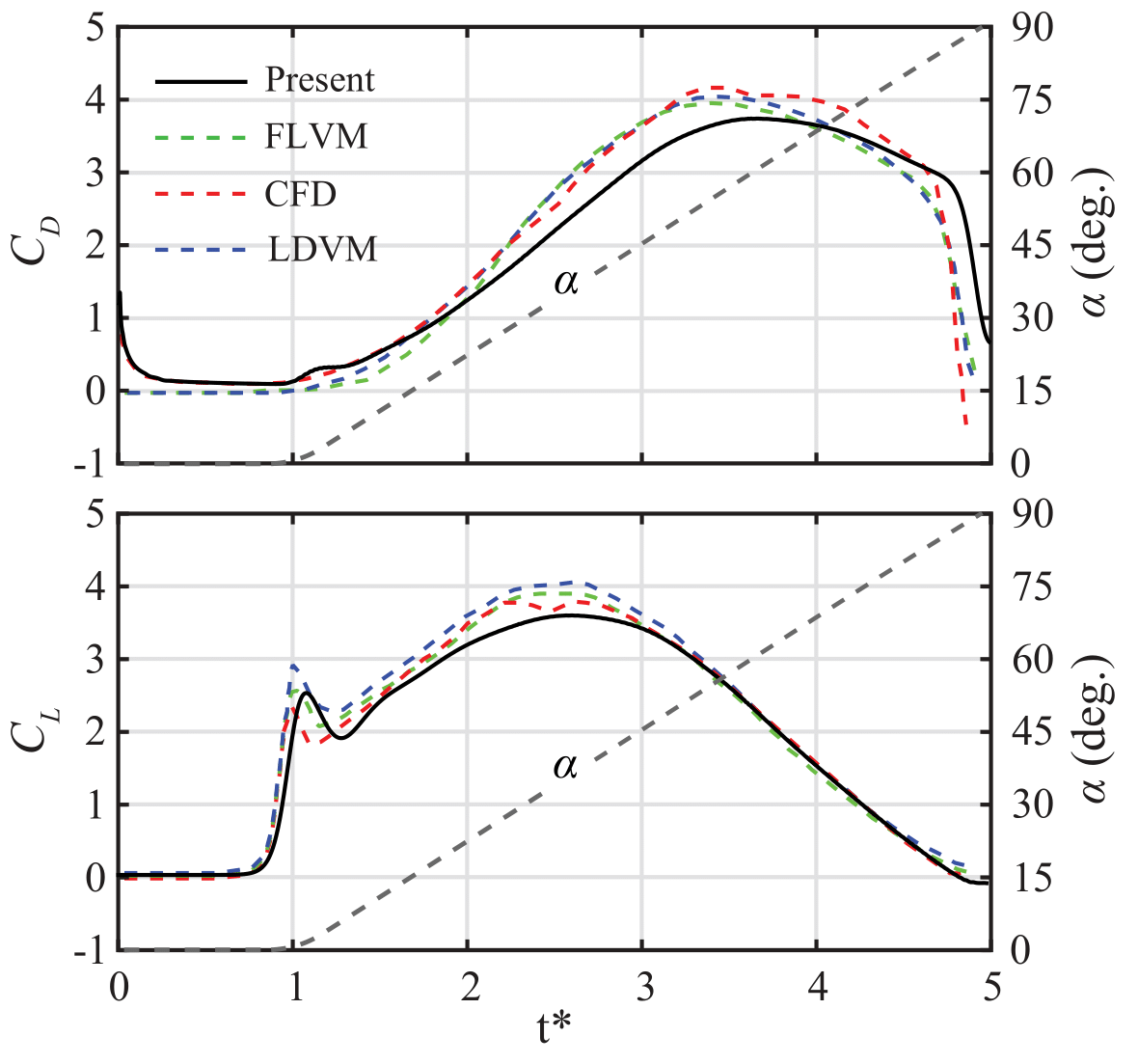}
    \caption{Variation of drag and lift coefficients with $t^*=tU_0/c$ from the present solver, FLVM, CFD, and LDVM. The scale for pitch-angle variation, $\alpha(t^*)$, is represented by the right-hand axes.}
    \label{fig.12}
\end{figure}

Figure \ref{fig.12} presents a comparison between the time history of the lift coefficient ($C_L$) and the drag coefficient ($C_D$) and the reference data for the evolution of the pitch angle. Although there is a similarity in the flow evolution, there is a noticeable phase difference between the hydrodynamic coefficients. Two processes can be observed in the lift coefficient ($C_L$): it increases with the pitch angle until it reaches the first positive critical value of $+2.6$ in approximately $1.2$ time units and a pitch angle of around $5^\circ$, which is consistent with the findings of Ramesh et al. \cite{ramesh2014discrete}. After this point, the lift coefficient ($C_L$) continues to increase steadily, reaching the second critical value by shedding clockwise vortices at each time step. These vortices form a leading-edge vortex (LEV) that detaches from the upper surface of the flat plate, and this LEV continues to grow until the end of the motion at $t^*=5.0$. In general, the lift and drag obtained from the SBoTFlow solver are in good agreement with the CFD results. However, the lift and drag coefficients indicate a time delay compared to the reference studies. Consequently, the magnitude coefficients at the latter critical point are lower compared to those of other studies.

\subsection{High-amplitude pitching airfoil}\label{v.5}
In this section, we investigate the effects of high-amplitude motion on airfoil geometries, specifically using the NACA 0018 airfoil at a Reynolds number of 1000. Our numerical computations focus on cases with periodic trajectories, and we base our kinematic model on the experimental investigation conducted by \=Otomo et. al. \cite{otomo2021unsteady}. To examine the accuracy of force predictions, we employ smoothed asymmetric triangular pitching kinematics, as shown in Fig. \ref{fig.14}. The computational domain and mesh configuration are similar to those used in the pitch-up maneuvered flat plate case study discussed in Sect. \ref{v.4}. This modified version of Theodorsen's theory \cite{theodorsen1935general} deviates from the original assumption of small sinusoidal oscillations, as analyzed in Appendix \ref{appendix.theodorsen}. The acceleration/deceleration time is denoted as $t_{\mathrm{a}}=0.15 T$, where $T=f_{\mathrm{p}}^{-1}$ represents the pitching period. Kinematics is described by a piecewise function, with the components of acceleration/deceleration represented by fourth-order polynomials.
\begin{equation}
    \alpha= \begin{cases}\dot{\alpha}_{1} t, & \left(0 \leq t<t_{1}\right), \\
    \frac{\dot{\alpha}_{1}}{2 t_{2}^{3}}\left(t-t_{2}\right)^{4}+\frac{\dot{\alpha}_{1}}{t_{2}^{2}}\left(t-t_{2}\right)^{3}+\alpha_{0}, & \left(t_{1} \leq t<t_{2}\right), \\
    -\frac{\dot{\alpha}_{2}}{2 t_{a}^{3}}\left(t-t_{2}\right)^{4}+\frac{\dot{\alpha}_{2}}{t_{a}^{2}}\left(t-t_{2}\right)^{3}+\alpha_{0}, & \left(t_{2} \leq t<t_{3}\right), \\
    \dot{\alpha}_{2}\left(t-t_{4}\right)-\frac{\dot{\alpha}_{2} t_{\mathrm{a}}}{2}-\alpha_{0}, & \left(t_{3} \leq t<t_{4}\right), \\
    \frac{\dot{\alpha}_{2}}{2 t_{2}^{3}}\left(t-t_{5}\right)^{4}+\frac{\dot{\alpha}_{2}}{t_{\mathrm{a}}^{2}}\left(t-t_{5}\right)^{3}-\alpha_{0}, & \left(t_{4} \leq t<t_{5}\right), \\
    -\frac{\dot{\alpha}_{1}}{2 t_{a}^{3}}\left(t-t_{5}\right)^{4}+\frac{\dot{\alpha}_{1}}{t_{\mathrm{a}}^{2}}\left(t-t_{5}\right)^{3}-\alpha_{0}, & \left(t_{5} \leq t<t_{6}\right), \\
    \dot{\alpha}_{1}(t-T), & \left(t_{6} \leq t<T\right),\end{cases}
\end{equation}
where $\dot{\alpha}_{1}$ and $\dot{\alpha}_{2}$ represent the rates of pitch in regions $0 \leq t \leq t_{1}$ and $t_{3} \leq t \leq t_{4}$, respectively, and are expressed as:
\begin{equation}
    \dot{\alpha}_{1}=\frac{2 \alpha_{0}}{\xi T-t_{\mathrm{a}}},
\end{equation}
\begin{equation}
    \dot{\alpha}_{2}=-\frac{2 \alpha_{0}}{(1-\xi) T-t_{\mathrm{a}}}.
\end{equation}

The degree of asymmetry is determined by the asymmetry parameter $\xi$, where the maximum angle of attack $\alpha_{0}$ occurs at $t_{2}=0.5 \xi$. We define the reduced frequency $k$ as $k=\pi f_{\mathrm{p}} c / U_0$, the pitching amplitude $\alpha_{0}$, and the asymmetry parameter $\xi$. We vary the reduced frequency $k=\pi f_{\mathrm{p}} c / U_{\infty}$, pitch amplitude $\alpha_{0}$, and asymmetry parameter $\xi$ so that $k \in\{0.22,0.44,0.66,0.88\}$, $\alpha_{0} \in\left\{4^{\circ}, 16^{\circ}, 64^{\circ}\right\}$, and $\xi=0.5$ for symmetric pitching motions.

\begin{figure}
    \centering
    \includegraphics[width=0.6\textwidth]{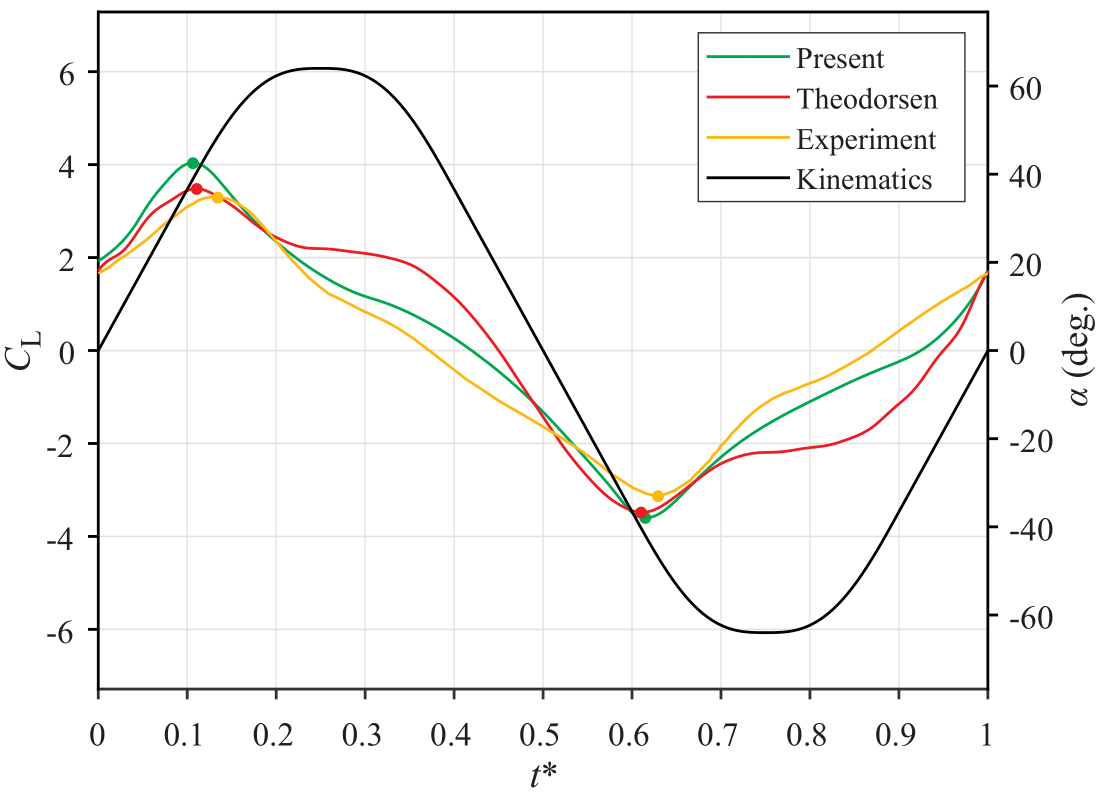}
    \caption{Time-history of the lift coefficient in a typical pitching period of the current SBoTFlow solver, Theodorsen's theory, and the experimental data by \=Otomo et. al. \cite{otomo2021unsteady} for $\alpha_0=64^\circ$, $k=0.22$, and $\xi=0.5$. The circles indicate the peaks of the lift coefficient. The black line represents the pitching motion referenced by the right vertical axis.}
    \label{fig.13}
\end{figure}

In order to analyze the flow topology, we examine a specific case where the pitch is symmetric with a value of $k=0.22$, $\alpha_{0}=64^{\circ}$, and $\xi=0.5$. This particular case demonstrates nonlinear effects on the forces due to the presence of shed vortices. Figure \ref{fig.13} shows the time-history of the lift coefficient in a typical pitching period obtained from the current SBoTFlow solver, Theodorsen's theory, and the experimental data by \=Otomo et. al. \cite{otomo2021unsteady} for $\alpha_0=64^\circ$, $k=0.22$, and $\xi=0.5$. Both the computation and the experiment agree well and show the same results throughout the typical pitching period. Although the maximum and minimum of the lift coefficient from computation have values closer to the theory than to the experiment, the theory prediction is overpredicted immediately after the pitch angle reaches maximum/minimum values. This overprediction is generated by the separation of the LEVs on the surface of the airfoil, forming a leading-edge suction force.

\begin{figure}
    \centering
    \includegraphics[width=0.8\textwidth]{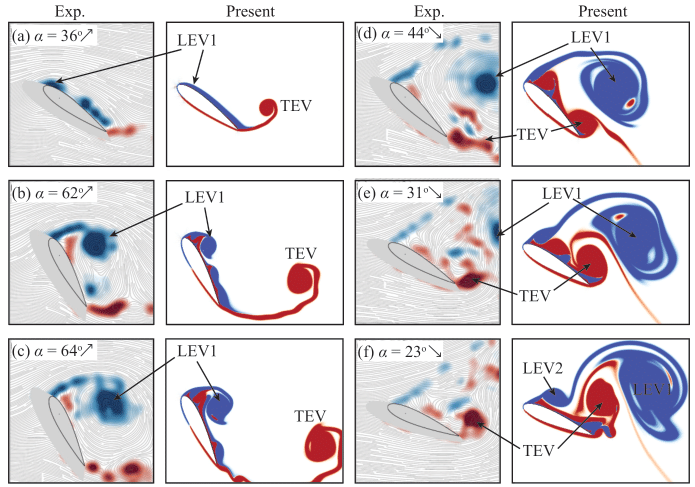}
    \caption{Identifed vortices in symmetric pitching case ($\alpha_0=64^\circ$, $k=0.22$, and $\xi=0.5$).}
    \label{fig.14}
\end{figure}

Figure \ref{fig.14} shows the identification of vortices generated by the pitch motion of the airfoil. To compare with the experimental findings of \=Otomo, we selected six specific moments in the first shed cycle to visualize the vorticity contour. The solver successfully captures the important vortices of leading-edge vortices (LEVs) and trailing-edge vortices (TEVs). The primary LEV (LEV1) is immediately followed by an increase in the angle of attack at $\alpha=36^\circ$, and its generation and growth in later stages can be observed. However, the secondary LEV (LEV2) was not captured in the experiments, but it was observed in detail through numerical computation, as shown in Fig. \ref{fig.14}f. On the other hand, both primary TEV (TEV1) and secondary TEV (TEV2) were clearly captured in both experiments and computations. The hairpin structure was observed in both cases before it broke down and generated TEV2.

\begin{figure}
    \centering
    \includegraphics[width=0.6\textwidth]{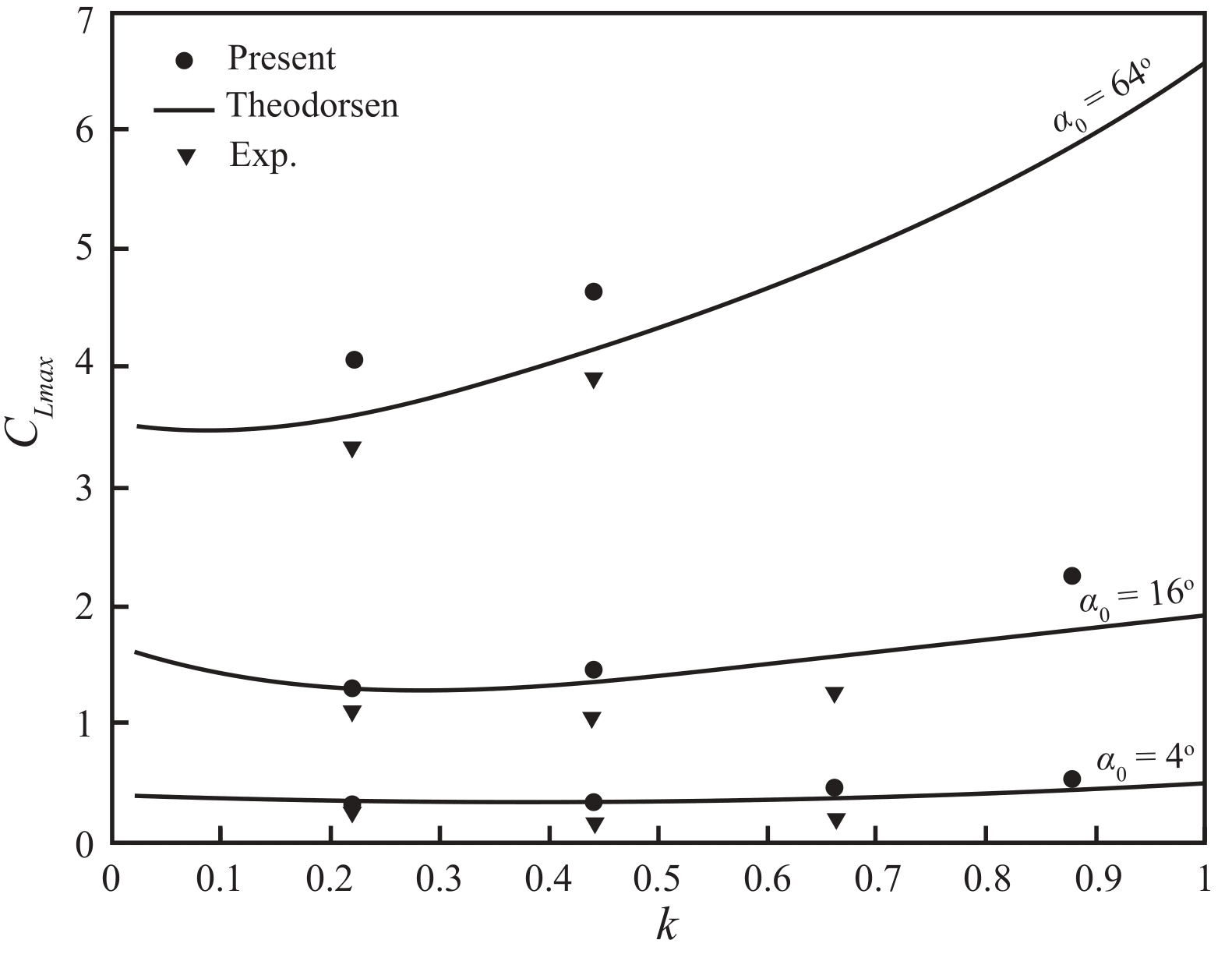}
    \caption{Maximum lift coefficient of symmetric pitching airfoils for different pitching amplitudes. Markers show the present results (circle) and experiment data (down triangle); solid lines indicate Theodorsen’s theory.}
    \label{fig.15}
\end{figure}

The collection of lift coefficient amplitudes is depicted in Fig. \ref{fig.15}. The figure compares the results obtained from the present solver, experimental data, and Theodorsen's theory prediction. In general, the theory tends to slightly overestimate the lift compared to the experimental data, while the numerical computation from the present solver underestimates it. Moreover, the largest differences in the amplitudes of the lift coefficient were observed in the cases with a pitching amplitude of $64^\circ$. On the other hand, for low-amplitude pitching cases like $4^\circ$ and $16^\circ$, the numerical computations of the present solver closely align with the theory prediction.

\subsection{High-frequency flapping wing}\label{v.6}
\begin{figure}
    \centering
    \includegraphics[width=0.85\textwidth]{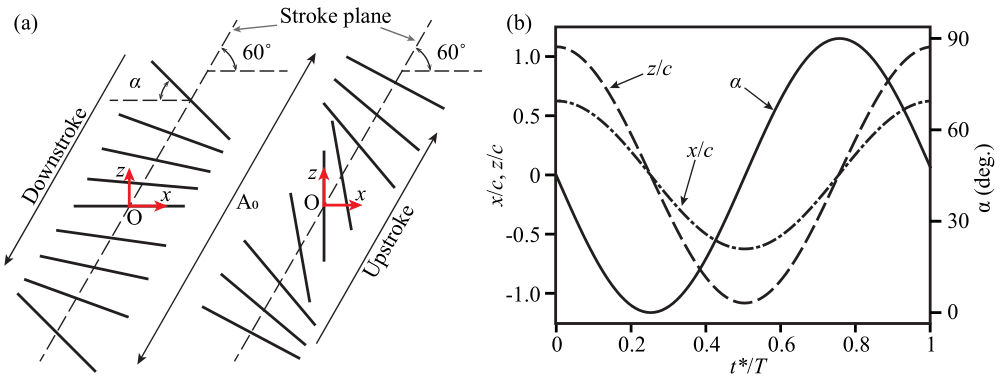}
    \caption{Illustration of flat-plate flapping motion: (a) sketch of kinematic of flapping motion; (b) evolution of the plate-center position and pitching angle in a flapping period.}
    \label{fig.16}
\end{figure}
In the last section, we analyze a conventional high-frequency flapping motion employing a two-dimensional flat plate configuration. The Reynolds number for this test is established at 157, as described by Wang et al.\cite{wang2000two}. The specific parameters utilized were derived from the published studies of Kim and Choi\cite{kim2007two} and Chen et al.\cite{chen2018unsteady}. In particular, the flat plate undergoes motion in a sloping stroke plane with an angle $\beta$ of $60^\circ$, replicating the kinematics of a dragonfly wing, as shown in Fig. \ref{fig.16}(a). The alteration in the pitching angle can be explained using the following equations:

% In the final section, we examine a standard high-frequency flapping motion using a 2D flat plate geometry. The Reynolds number for this test is set to 157, as documented by Wang et al.\cite{wang2000two}. The specific parameters used were obtained from the published works of Kim and Choi\cite{kim2007two} and Chen et al.\cite{chen2018unsteady}. Specifically, the flat plate moves in a sloped stroke plane at an angle $\beta$ of $60^\circ$, imitating the motion kinematics of a dragonfly wing as depicted in Fig. \ref{fig.15}(a). The change in pitching angle can be described by the following equations:
\begin{equation}
    \alpha(t)=\alpha_{0}-\frac{\pi}{4} \sin \left(\frac{2 \pi t}{T}+\varphi\right),
\end{equation}
where $\alpha_0$ is average pitching angle, $T$ is the period of a flapping cycle, and $\varphi$ is the phase difference.  The center position of the flat plate is given by
\begin{align}
    x_c=\frac{A_0}{2}\mathrm{cos}\left(\frac{2\pi t^*}{T}\right)\mathrm{cos}(\beta),\\
    y_c=\frac{A_0}{2}\mathrm{cos}\left(\frac{2\pi t^*}{T}\right)\mathrm{sin}(\beta),
\end{align}
where $A_0$ denotes the magnitude of the vertical displacement.
\begin{figure}
    \centering
    \includegraphics[width=0.4\textwidth]{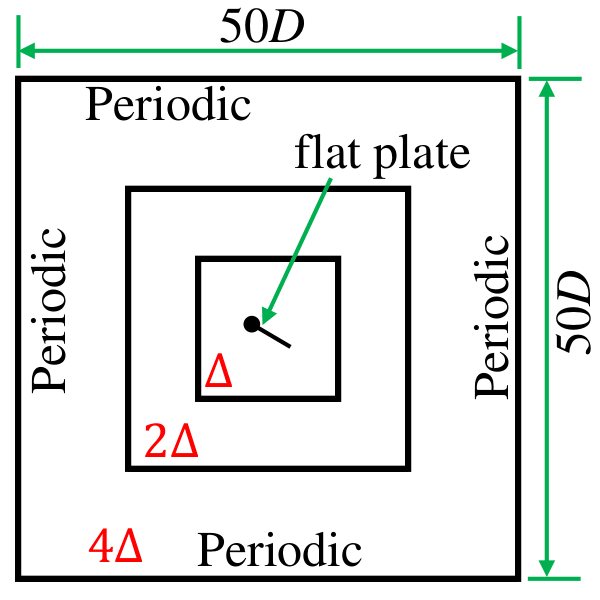}
    \caption{Computational domain of flapping wing simulation.}
    \label{fig.17}
\end{figure}

Figure \ref{fig.16}(b) depicts the movement of the wings throughout a single flapping cycle and the angle of displacement in pitch. The forward displacement is represented by the variable $x$, whereas the downward displacement is represented by the variable $z$. The investigation employs the following parameters: a flat plate with a chord length of $c=1$ and a thickness of 4\%, a heaving amplitude of $A_0=2.5c$, an average pitch angle of $\alpha_0$, a phase difference of $\phi=0$, and a flapping period of $T=2.5\pi$. This case study employs a flow configuration featuring a square domain size of $50D\times50D$ and period boundary conditions on every side. Furthermore, the three refinement stages mentioned in Sects. \ref{v.4} and \ref{v.5} are employed to guarantee a domain size of $10D\times10D$ with the best grid resolution focused around the flat plate, as shown in Fig. \ref{fig.17}. The SBoTFlow solver's instantaneous force produced a steady periodic oscillation across two flapping cycles by adhering to the established motion kinematics. 

% with $A_0$ representing the amplitude of heaving. Figure \ref{fig.15}(b) illustrates the displacement of the wings during one flapping cycle and the displacement angle in pitch. The displacements in the forward and downward directions are denoted by $x$ and $z$ respectively. In this study, the following parameters are used: flat plate chord $c=1$ with thickness 4\%, heaving amplitude $A_0=2.5c$, average pitch angle $\alpha_0$, phase difference $\phi=0$, and flapping period $T=2.5\pi$. By following the prescribed motion kinematics, the SBoTFlow solver's instantaneous force achieved a stable periodic oscillation over two flapping cycles. 

\begin{figure}
    \centering
    \includegraphics[width=0.9\textwidth]{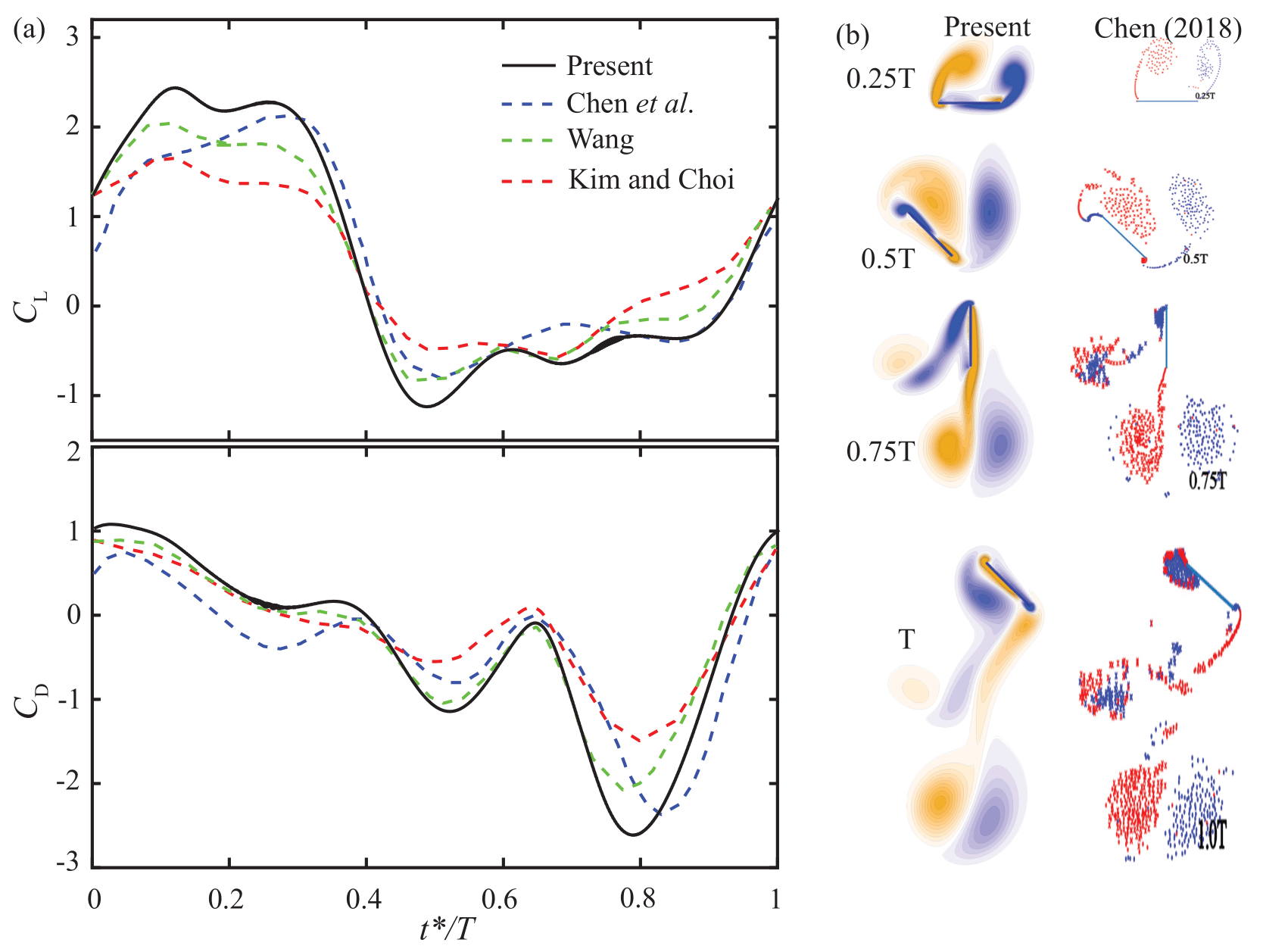}
    \caption{Comparison of the lift and drag coefficients in a flapping period (a); Progression of vorticity field during one flapping period from the SBoTFlow solver and CFD by Chen et al.\cite{chen2018unsteady} (b)}
    \label{fig.18}
\end{figure}

Figure \ref{fig.18}(a) compares the results of the lift and drag coefficients obtained from the SBoTFlow solver with those from previous studies. The CFD-based numerical method of Kim and Choi \cite{kim2007two} and Wang \cite{wang2000two}, as well as the theoretical method of Chen et al. \cite{chen2018unsteady}, were used for comparison. Variations in the lift and drag coefficients are similar to those observed in the computational results from the mentioned references. Both the downstroke and the upstroke exhibit two peak values that are in good agreement with the CFD methods. However, the theoretical method only confirms the second peak, which is caused by the influence of fluid viscosity at lower Reynolds numbers \cite{kamisawa2008optimum}. The mean lift coefficient obtained from the present SBoTFlow solver, $C_L=0.6$, is higher than the values obtained from the works of Wang ($0.49$) and Chen ($0.42$), while the variability of the drag coefficient during the flapping period based on SBoTFlow shows a similar trend to the CFD result of Wang. Consequently, the mean drag coefficient, $C_D=-0.34$, is relatively close to the CFD result of Wang ($C_{D}=-0.28$), but significantly different from the other CFD result of Kim and Choi ($C_{D}=-0.5$).

Figure \ref{fig.18}(b) illustrates the progression of the flow field during the flapping period. The vortices at the leading edge (LEV) and trailing edge (TEV) are divided into pairs. As the wings move, a pair of vortices are formed at each edge in opposite rotational directions. The clockwise vortices are represented by red dots and yellow contours, while the anticlockwise vortices are indicated by blue dots and blue contours. Wing rotation combines these vortices into a dipole. Due to the influence of each vortex on the other, they form a co-moving pair. The dipole descends, carrying momentum and generating lift on the wing. The self-induced flow removes the vortices from the wing, preventing interference with the vortices in the subsequent cycle. In this case, the shedding frequency matches the flapping frequency. These findings are consistent with those reported by Wang \cite{wang2000two}.
% \begin{figure}
%     \centering
%     \includegraphics[width=0.4\textwidth]{figure19.png}
%     \caption{Progression of vorticity field during one flapping period from the SBoTFlow solver (left-hand column) and CFD by Chen et al.\cite{chen2018unsteady} (right-hand column)}
%     \label{fig.19}
% \end{figure}
\section{Conclusions} \label{c}
We presented a scalable computational framework named SBoTFlow that utilizes the lattice Boltzmann method, the multidirect forcing immersed boundary scheme, and topology-confined block refinement for simulations involving moving bodies. This framework offers a versatile and automated tool for generating meshes tailored for parallel computation in high-performance computing (HPC) environments. We validated and analyzed the effectiveness of our approach by conducting studies on different Reynolds numbers, which encompassed both stationary and moving-body settings. The results demonstrate outstanding performance with an improvement in memory usage as compared to typical lattice Boltzmann implementations. The calculated results exhibited exceptional concurrence with experimental and computational data obtained from prior investigations, thus affirming the dependability of our simulations for both stationary and moving objects. Specifically, our solver showed a strong predictive capacity for wing lift during pitching movements with high amplitudes of up to $64^\circ$. It successfully depicted the non-linear characteristics of the fluid in these circumstances. Our solver recorded higher magnitudes of lift and drag for the flapping motion, particularly at high frequencies of $f=1/2.5\pi$, as compared to the references. Nevertheless, it is crucial to acknowledge that the Reynolds numbers employed in this investigation were rather limited, amounting to less than 1000. In summary, our paradigm is uncomplicated and its inherent parallelism indicates prospective uses in the analysis of agile movements, such as insect flying or wing aerodynamics. Subsequent research will prioritize investigating the feasibility of expanding the single- and multi-GPU designs to accommodate extensive simulations that require the computational capabilities of current pre-exascale supercomputers and future exascale infrastructures.

\appendix

\section{Theodorsen's theory for high amplitude} \label{appendix.theodorsen}
The classical theories proposed by Theodorsen \cite{theodorsen1935general} and von K\'arm\'an and Sears \cite{thomson1883treatise} focused on studying the small sinusoidal oscillation of a flat-plate. They employed linear potential flow in attached flow conditions to analyze this phenomenon. Several experimental investigations \cite{kang2009fluid,mcgowan2011investigations,baik2012unsteady,liu2015unsteady,cordes2017note} were carried out under these conditions, and the theoretical analysis showed good agreement with the experimental results. The study examined a wide range of angles of attack ($\alpha$), with a maximum value of $64^\circ$, resulting in significant flow separation. This violated the expected limit of the theory's applicability. According to Theodorsen's theory, the pressure difference between the upper and lower surfaces of the flat plate contributes to the normal forces, which can be expressed as the sum of non-circulatory (also known as apparent mass in the literature) and circulatory terms.
\begin{equation}
    \begin{aligned}
    C_{\mathrm{N}}^{\mathrm{th}}= & \underbrace{\frac{\pi c}{2 U_{0}^2}\left[\dot{\alpha} U_{0}-\frac{c}{2} \ddot{\alpha}\left(2 x_{\mathrm{p}}-1\right)\right]}_{\text {non-circulatory }} 
     +\underbrace{\frac{2 \pi C(k)}{U_{\infty}}\left[\alpha U_{0}+\frac{c}{2} \dot{\alpha}\left(\frac{3}{2}-2 x_{\mathrm{p}}\right)\right]}_{\text {circulatory }}.
    \end{aligned}
\end{equation}

The equation for computing the lift coefficient is given by: 
\begin{equation}
C_{\mathrm{L}}^{\mathrm{th}}=C_{\mathrm{N}}^{\mathrm{th}} \cos \alpha,
\end{equation} 
where $\dot{\alpha}$ and $\ddot{\alpha}$ represent the first-order and second-order derivatives of the angle of attack in pitching motion, and $x_{\mathrm{p}}$ is the normalized location of the pitching axis from the leading edge with respect to the chord line $c$. The equation $C(k)$ is formulated as: 
\begin{equation}
C(k)=\frac{H_1^{(2)}(k)}{H_1^{(2)}(k)+i H_0^{(2)}(k)},
\end{equation}
where $H_n^{(2)}$ represents the Hankel function of the second kind of order $n$. The Theodorsen function $C(k)$ determines the influence of the wake on the strength of the vortex sheet along the surface of an airfoil. It is important to note that Theodorsen's theory is applicable only to sinusoidal oscillations. To address this limitation, we expanded the smoothed triangular kinematics using the first twenty Fourier harmonics. This expansion resulted in curves that differed from the original kinematics by less than $1\%$. Finally, the total lift coefficient is computed by summing the lift coefficients calculated for each of the 20 Fourier harmonics.
\section*{Acknowledgments}
We are deeply grateful to all anonymous reviewers for their invaluable, detailed, and informative suggestions and comments on the drafts of this article.

\section*{Disclosure statement}
No potential conflict of interest was reported by the author(s).

% \section*{Funding}

\bibliography{SBoTFlow.bib}

\begin{thebibliography}{10}

\bibitem{sane2003aerodynamics}
S.~P. Sane, ``The aerodynamics of insect flight,'' {\em Journal of experimental biology}, vol.~206, no.~23, pp.~4191--4208, 2003.

\bibitem{nakata2012fluid}
T.~Nakata and H.~Liu, ``A fluid--structure interaction model of insect flight with flexible wings,'' {\em Journal of Computational Physics}, vol.~231, no.~4, pp.~1822--1847, 2012.

\bibitem{abdelmaksoud2020control}
S.~I. Abdelmaksoud, M.~Mailah, and A.~M. Abdallah, ``Control strategies and novel techniques for autonomous rotorcraft unmanned aerial vehicles: A review,'' {\em IEEE Access}, vol.~8, pp.~195142--195169, 2020.

\bibitem{suzuki2019effect}
K.~Suzuki, T.~Aoki, and M.~Yoshino, ``Effect of chordwise wing flexibility on flapping flight of a butterfly model using immersed-boundary lattice boltzmann simulations,'' {\em Physical Review E}, vol.~100, no.~1, p.~013104, 2019.

\bibitem{huang2019recent}
W.-X. Huang and F.-B. Tian, ``Recent trends and progress in the immersed boundary method,'' {\em Proceedings of the Institution of Mechanical Engineers, Part C: Journal of Mechanical Engineering Science}, vol.~233, no.~23-24, pp.~7617--7636, 2019.

\bibitem{majumdar2020capturing}
D.~Majumdar, C.~Bose, and S.~Sarkar, ``Capturing the dynamical transitions in the flow-field of a flapping foil using immersed boundary method,'' {\em Journal of Fluids and Structures}, vol.~95, p.~102999, 2020.

\bibitem{tao2022lattice}
S.~Tao, L.~Wang, Q.~He, J.~Chen, and J.~Luo, ``Lattice boltzmann simulation of complex thermal flows via a simplified immersed boundary method,'' {\em Journal of Computational Science}, vol.~65, p.~101878, 2022.

\bibitem{greenblatt2022flow}
D.~Greenblatt and D.~R. Williams, ``Flow control for unmanned air vehicles,'' {\em Annual Review of Fluid Mechanics}, vol.~54, pp.~383--412, 2022.

\bibitem{xiao2013flow}
Q.~Xiao, W.~Liu, and A.~Incecik, ``Flow control for vatt by fixed and oscillating flap,'' {\em Renewable Energy}, vol.~51, pp.~141--152, 2013.

\bibitem{kim2001immersed}
J.~Kim, D.~Kim, and H.~Choi, ``An immersed-boundary finite-volume method for simulations of flow in complex geometries,'' {\em Journal of computational physics}, vol.~171, no.~1, pp.~132--150, 2001.

\bibitem{wang2015immersed}
Y.~Wang, C.~Shu, C.~Teo, and J.~Wu, ``An immersed boundary-lattice boltzmann flux solver and its applications to fluid--structure interaction problems,'' {\em Journal of Fluids and Structures}, vol.~54, pp.~440--465, 2015.

\bibitem{kim2019immersed}
W.~Kim and H.~Choi, ``Immersed boundary methods for fluid-structure interaction: A review,'' {\em International Journal of Heat and Fluid Flow}, vol.~75, pp.~301--309, 2019.

\bibitem{tseng2003ghost}
Y.-H. Tseng and J.~H. Ferziger, ``A ghost-cell immersed boundary method for flow in complex geometry,'' {\em Journal of computational physics}, vol.~192, no.~2, pp.~593--623, 2003.

\bibitem{terashima2009front}
H.~Terashima and G.~Tryggvason, ``A front-tracking/ghost-fluid method for fluid interfaces in compressible flows,'' {\em Journal of Computational Physics}, vol.~228, no.~11, pp.~4012--4037, 2009.

\bibitem{peskin1972flow}
C.~S. Peskin, {\em Flow patterns around heart valves: a digital computer method for solving the equations of motion}.
\newblock Yeshiva University, 1972.

\bibitem{peskin1977numerical}
C.~S. Peskin, ``Numerical analysis of blood flow in the heart,'' {\em Journal of computational physics}, vol.~25, no.~3, pp.~220--252, 1977.

\bibitem{nishiguchi2019full}
K.~Nishiguchi, R.~Bale, S.~Okazawa, and M.~Tsubokura, ``Full eulerian deformable solid-fluid interaction scheme based on building-cube method for large-scale parallel computing,'' {\em International Journal for Numerical Methods in Engineering}, vol.~117, no.~2, pp.~221--248, 2019.

\bibitem{kruger2017lattice}
T.~Kr{\"u}ger, H.~Kusumaatmaja, A.~Kuzmin, O.~Shardt, G.~Silva, and E.~M. Viggen, ``The lattice boltzmann method,'' {\em Springer International Publishing}, vol.~10, no.~978-3, pp.~4--15, 2017.

\bibitem{chen1998lattice}
S.~Chen and G.~D. Doolen, ``Lattice boltzmann method for fluid flows,'' {\em Annual review of fluid mechanics}, vol.~30, no.~1, pp.~329--364, 1998.

\bibitem{yeomans2006mesoscale}
J.~Yeomans, ``Mesoscale simulations: Lattice boltzmann and particle algorithms,'' {\em Physica A: Statistical Mechanics and its Applications}, vol.~369, no.~1, pp.~159--184, 2006.

\bibitem{lu2022analyses}
J.~Lu, H.~Lei, C.~Dai, L.~Yang, and C.~Shu, ``Analyses and reconstruction of the lattice boltzmann flux solver,'' {\em Journal of Computational Physics}, vol.~453, p.~110923, 2022.

\bibitem{wang2023recent}
L.~Wang, Z.~Liu, and M.~Rajamuni, ``Recent progress of lattice boltzmann method and its applications in fluid-structure interaction,'' {\em Proceedings of the Institution of Mechanical Engineers, Part C: Journal of Mechanical Engineering Science}, vol.~237, no.~11, pp.~2461--2484, 2023.

\bibitem{he1997priori}
X.~He and L.-S. Luo, ``A priori derivation of the lattice boltzmann equation,'' {\em Physical Review E}, vol.~55, no.~6, p.~R6333, 1997.

\bibitem{succi2018lattice}
S.~Succi and S.~Succi, {\em The lattice Boltzmann equation: for complex states of flowing matter}.
\newblock Oxford university press, 2018.

\bibitem{bhatnagar1954model}
P.~L. Bhatnagar, E.~P. Gross, and M.~Krook, ``A model for collision processes in gases. i. small amplitude processes in charged and neutral one-component systems,'' {\em Physical review}, vol.~94, no.~3, p.~511, 1954.

\bibitem{he1997theory}
X.~He and L.-S. Luo, ``Theory of the lattice boltzmann method: From the boltzmann equation to the lattice boltzmann equation,'' {\em Physical review E}, vol.~56, no.~6, p.~6811, 1997.

\bibitem{ou2022directional}
Z.~Ou, C.~Chi, L.~Guo, and D.~Th{\'e}venin, ``A directional ghost-cell immersed boundary method for low mach number reacting flows with interphase heat and mass transfer,'' {\em Journal of Computational Physics}, vol.~468, p.~111447, 2022.

\bibitem{liu2023investigation}
W.-T. Liu, A.-M. Zhang, X.-H. Miao, F.-R. Ming, and Y.-L. Liu, ``Investigation of hydrodynamics of water impact and tail slamming of high-speed water entry with a novel immersed boundary method,'' {\em Journal of Fluid Mechanics}, vol.~958, p.~A42, 2023.

\bibitem{peskin1996case}
C.~Peskin, D.~McQueen, and H.~Othmer, ``Case studies in mathematical modeling: ecology, physiology, and cell biology,'' 1996.

\bibitem{peskin2002immersed}
C.~S. Peskin, ``The immersed boundary method,'' {\em Acta numerica}, vol.~11, pp.~479--517, 2002.

\bibitem{beyer1992analysis}
R.~P. Beyer and R.~J. LeVeque, ``Analysis of a one-dimensional model for the immersed boundary method,'' {\em SIAM Journal on Numerical Analysis}, vol.~29, no.~2, pp.~332--364, 1992.

\bibitem{goldstein1993modeling}
D.~Goldstein, R.~Handler, and L.~Sirovich, ``Modeling a no-slip flow boundary with an external force field,'' {\em Journal of computational physics}, vol.~105, no.~2, pp.~354--366, 1993.

\bibitem{breugem2012second}
W.-P. Breugem, ``A second-order accurate immersed boundary method for fully resolved simulations of particle-laden flows,'' {\em Journal of Computational Physics}, vol.~231, no.~13, pp.~4469--4498, 2012.

\bibitem{kang2011direct}
S.~K. Kang and Y.~A. Hassan, ``A direct-forcing immersed boundary method for the thermal lattice boltzmann method,'' {\em Computers \& Fluids}, vol.~49, no.~1, pp.~36--45, 2011.

\bibitem{thorimbert2018lattice}
Y.~Thorimbert, F.~Marson, A.~Parmigiani, B.~Chopard, and J.~L{\"a}tt, ``Lattice boltzmann simulation of dense rigid spherical particle suspensions using immersed boundary method,'' {\em Computers \& Fluids}, vol.~166, pp.~286--294, 2018.

\bibitem{cheylan2021immersed}
I.~Cheylan, J.~Favier, and P.~Sagaut, ``Immersed boundary conditions for moving objects in turbulent flows with the lattice-boltzmann method,'' {\em Physics of Fluids}, vol.~33, no.~9, 2021.

\bibitem{feng2004immersed}
Z.-G. Feng and E.~E. Michaelides, ``The immersed boundary-lattice boltzmann method for solving fluid--particles interaction problems,'' {\em Journal of computational physics}, vol.~195, no.~2, pp.~602--628, 2004.

\bibitem{kang2011comparative}
S.~K. Kang and Y.~A. Hassan, ``A comparative study of direct-forcing immersed boundary-lattice boltzmann methods for stationary complex boundaries,'' {\em International Journal for Numerical Methods in Fluids}, vol.~66, no.~9, pp.~1132--1158, 2011.

\bibitem{shyy2016aerodynamics}
W.~Shyy, C.-k. Kang, P.~Chirarattananon, S.~Ravi, and H.~Liu, ``Aerodynamics, sensing and control of insect-scale flapping-wing flight,'' {\em Proceedings of the Royal Society A: Mathematical, Physical and Engineering Sciences}, vol.~472, no.~2186, p.~20150712, 2016.

\bibitem{tian2011efficient}
F.-B. Tian, H.~Luo, L.~Zhu, J.~C. Liao, and X.-Y. Lu, ``An efficient immersed boundary-lattice boltzmann method for the hydrodynamic interaction of elastic filaments,'' {\em Journal of computational physics}, vol.~230, no.~19, pp.~7266--7283, 2011.

\bibitem{zhu2021numerical}
Y.~Zhu, F.-B. Tian, J.~Young, J.~C. Liao, and J.~C. Lai, ``A numerical study of fish adaption behaviors in complex environments with a deep reinforcement learning and immersed boundary--lattice boltzmann method,'' {\em Scientific Reports}, vol.~11, no.~1, p.~1691, 2021.

\bibitem{moin1998direct}
P.~Moin and K.~Mahesh, ``Direct numerical simulation: a tool in turbulence research,'' {\em Annual review of fluid mechanics}, vol.~30, no.~1, pp.~539--578, 1998.

\bibitem{cao2022topological}
Y.~Cao, T.~Tamura, D.~Zhou, Y.~Bao, and Z.~Han, ``Topological description of near-wall flows around a surface-mounted square cylinder at high reynolds numbers,'' {\em Journal of Fluid Mechanics}, vol.~933, p.~A39, 2022.

\bibitem{nakahashi2004building}
K.~Nakahashi and L.~Kim, ``Building-cube method for large-scale, high resolution flow computations,'' in {\em 42nd AIAA Aerospace Sciences Meeting and Exhibit}, p.~434, 2004.

\bibitem{nakahashi2006three}
K.~Nakahashi, A.~Kitoh, Y.~Sakurai, and M.~Meinke, ``Three-dimensional flow computations around an airfoil by building-cube method,'' in {\em 44th AIAA Aerospace Sciences Meeting and Exhibit}, p.~1104, 2006.

\bibitem{ishida2008efficient}
T.~Ishida, S.~Takahashi, and K.~Nakahashi, ``Efficient and robust cartesian mesh generation for building-cube method,'' {\em Journal of Computational Science and Technology}, vol.~2, no.~4, pp.~435--446, 2008.

\bibitem{duong2022low}
V.~D. Duong, V.~D. Nguyen, V.~T. Nguyen, and I.~L. Ngo, ``Low-reynolds-number wake of three tandem elliptic cylinders,'' {\em Physics of Fluids}, vol.~34, no.~4, 2022.

\bibitem{duong2024near}
V.~D. Duong, V.~L. Nguyen, V.~T. Nguyen, P.~S. Palar, L.~R. Zuhal, A.~T. Le, J.-K. Lin, and W.-C. Wang, ``Near-moving-wall flows past three tandem elliptical cylinders at low reynolds number of 150,'' {\em Physics of Fluids}, vol.~36, no.~1, 2024.

\bibitem{hafen2023numerical}
N.~Hafen, J.~R. Thieringer, J.~Meyer, M.~J. Krause, and A.~Dittler, ``Numerical investigation of detachment and transport of particulate structures in wall-flow filters using lattice boltzmann methods,'' {\em Journal of Fluid Mechanics}, vol.~956, p.~A30, 2023.

\bibitem{guo2002lattice}
Z.~Guo and T.~Zhao, ``Lattice boltzmann model for incompressible flows through porous media,'' {\em Physical review E}, vol.~66, no.~3, p.~036304, 2002.

\bibitem{shu2014development}
C.~Shu, Y.~Wang, C.~Teo, and J.~Wu, ``Development of lattice boltzmann flux solver for simulation of incompressible flows,'' {\em Advances in Applied Mathematics and Mechanics}, vol.~6, no.~4, pp.~436--460, 2014.

\bibitem{du2006multi}
R.~Du, B.~Shi, and X.~Chen, ``Multi-relaxation-time lattice boltzmann model for incompressible flow,'' {\em Physics Letters A}, vol.~359, no.~6, pp.~564--572, 2006.

\bibitem{malaspinas2014wall}
O.~Malaspinas and P.~Sagaut, ``Wall model for large-eddy simulation based on the lattice boltzmann method,'' {\em Journal of Computational Physics}, vol.~275, pp.~25--40, 2014.

\bibitem{kamatsuchi2007turbulent}
T.~Kamatsuchi, ``Turbulent flow simulation around complex geometries with cartesian grid method,'' in {\em 45th AIAA Aerospace Sciences Meeting and Exhibit}, p.~1459, 2007.

\bibitem{bader2012space}
M.~Bader, {\em Space-filling curves: an introduction with applications in scientific computing}, vol.~9.
\newblock Springer Science \& Business Media, 2012.

\bibitem{rohde2006generic}
M.~Rohde, D.~Kandhai, J.~J. Derksen, and H.~E. Van~den Akker, ``A generic, mass conservative local grid refinement technique for lattice-boltzmann schemes,'' {\em International journal for numerical methods in fluids}, vol.~51, no.~4, pp.~439--468, 2006.

\bibitem{schornbaum2018block}
F.~Schornbaum, {\em Block-structured adaptive mesh refinement for simulations on extreme-scale supercomputers}.
\newblock PhD thesis, Friedrich-Alexander-Universit{\"a}t Erlangen-N{\"u}rnberg (FAU), 2018.

\bibitem{vtkBook}
W.~Schroeder, K.~Martin, and B.~Lorensen, {\em The Visualization Toolkit (4th ed.)}.
\newblock Kitware, 2006.

\bibitem{taylor1937mechanism}
G.~I. Taylor and A.~E. Green, ``Mechanism of the production of small eddies from large ones,'' {\em Proceedings of the Royal Society of London. Series A-Mathematical and Physical Sciences}, vol.~158, no.~895, pp.~499--521, 1937.

\bibitem{cai2017moving}
S.-G. Cai, A.~Ouahsine, J.~Favier, and Y.~Hoarau, ``Moving immersed boundary method,'' {\em International Journal for Numerical Methods in Fluids}, vol.~85, no.~5, pp.~288--323, 2017.

\bibitem{rajan2021flow}
I.~Rajan and D.~A. Perumal, ``Flow dynamics of lid-driven cavities with obstacles of various shapes and configurations using the lattice boltzmann method,'' {\em Journal of Thermal Engineering}, vol.~7, no.~2, pp.~83--102, 2021.

\bibitem{ghias2004non}
R.~Ghias, R.~Mittal, and T.~Lund, ``A non-body conformal grid method for simulation of compressible flows with complex immersed boundaries,'' in {\em 42nd AIAA aerospace sciences meeting and exhibit}, p.~80, 2004.

\bibitem{de2006immersed}
P.~De~Palma, M.~De~Tullio, G.~Pascazio, and M.~Napolitano, ``An immersed-boundary method for compressible viscous flows,'' {\em Computers \& fluids}, vol.~35, no.~7, pp.~693--702, 2006.

\bibitem{rajani2009numerical}
B.~Rajani, A.~Kandasamy, and S.~Majumdar, ``Numerical simulation of laminar flow past a circular cylinder,'' {\em Applied Mathematical Modelling}, vol.~33, no.~3, pp.~1228--1247, 2009.

\bibitem{boukharfane2018combined}
R.~Boukharfane, F.~H.~E. Ribeiro, Z.~Bouali, and A.~Mura, ``A combined ghost-point-forcing/direct-forcing immersed boundary method (ibm) for compressible flow simulations,'' {\em Computers \& fluids}, vol.~162, pp.~91--112, 2018.

\bibitem{de2020sharp}
F.~De~Vanna, F.~Picano, and E.~Benini, ``A sharp-interface immersed boundary method for moving objects in compressible viscous flows,'' {\em Computers \& Fluids}, vol.~201, p.~104415, 2020.

\bibitem{bouard1980early}
R.~Bouard and M.~Coutanceau, ``The early stage of development of the wake behind an impulsively started cylinder for 40< re< 104,'' {\em Journal of Fluid Mechanics}, vol.~101, no.~3, pp.~583--607, 1980.

\bibitem{duong2021fluid}
V.~D. Duong, L.~R. Zuhal, and H.~Muhammad, ``Fluid--structure coupling in time domain for dynamic stall using purely lagrangian vortex method,'' {\em CEAS Aeronautical Journal}, vol.~12, pp.~381--399, 2021.

\bibitem{kim2012vortex}
Y.-C. Kim, J.-C. Suh, and K.-J. Lee, ``Vortex-in-cell method combined with a boundary element method for incompressible viscous flow analysis,'' {\em International Journal for Numerical Methods in Fluids}, vol.~69, no.~10, pp.~1567--1583, 2012.

\bibitem{taneda1956experimental}
S.~Taneda, ``Experimental investigation of the wake behind a sphere at low reynolds numbers,'' {\em Journal of the physical society of Japan}, vol.~11, no.~10, pp.~1104--1108, 1956.

\bibitem{ramesh2014discrete}
K.~Ramesh, A.~Gopalarathnam, K.~Granlund, M.~V. Ol, and J.~R. Edwards, ``Discrete-vortex method with novel shedding criterion for unsteady aerofoil flows with intermittent leading-edge vortex shedding,'' {\em Journal of Fluid Mechanics}, vol.~751, pp.~500--538, 2014.

\bibitem{eldredge2007numerical}
J.~D. Eldredge, ``Numerical simulation of the fluid dynamics of 2d rigid body motion with the vortex particle method,'' {\em Journal of Computational Physics}, vol.~221, no.~2, pp.~626--648, 2007.

\bibitem{eldredge2009computational}
J.~Eldredge, C.~Wang, and M.~Ol, ``A computational study of a canonical pitch-up, pitch-down wing maneuver,'' in {\em 39th AIAA fluid dynamics conference}, p.~3687, 2009.

\bibitem{wang2013low}
C.~Wang and J.~D. Eldredge, ``Low-order phenomenological modeling of leading-edge vortex formation,'' {\em Theoretical and Computational Fluid Dynamics}, vol.~27, pp.~577--598, 2013.

\bibitem{otomo2021unsteady}
S.~{\=O}tomo, S.~Henne, K.~Mulleners, K.~Ramesh, and I.~M. Viola, ``Unsteady lift on a high-amplitude pitching aerofoil,'' {\em Experiments in Fluids}, vol.~62, pp.~1--18, 2021.

\bibitem{theodorsen1935general}
T.~Theodorsen, ``General theory of aerodynamic instability and the mechanism of,'' {\em Annual Report of the National Advisory Committee for Aeronautics}, vol.~268, p.~413, 1935.

\bibitem{wang2000two}
Z.~J. Wang, ``Two dimensional mechanism for insect hovering,'' {\em Physical review letters}, vol.~85, no.~10, p.~2216, 2000.

\bibitem{kim2007two}
D.~Kim and H.~Choi, ``Two-dimensional mechanism of hovering flight by single flapping wing,'' {\em Journal of Mechanical Science and Technology}, vol.~21, pp.~207--221, 2007.

\bibitem{chen2018unsteady}
S.~Chen, H.~Li, S.~Guo, M.~Tong, and B.~Ji, ``Unsteady aerodynamic model of flexible flapping wing,'' {\em Aerospace Science and Technology}, vol.~80, pp.~354--367, 2018.

\bibitem{kamisawa2008optimum}
Y.~Kamisawa and K.~Isogai, ``Optimum flapping wing motions of dragonfly,'' {\em Transactions of the Japan Society for Aeronautical and Space Sciences}, vol.~51, no.~172, pp.~114--123, 2008.

\bibitem{thomson1883treatise}
J.~J. Thomson, {\em A Treatise on the Motion of Vortex Rings: an essay to which the Adams prize was adjudged in 1882, in the University of Cambridge}.
\newblock Macmillan, 1883.

\bibitem{kang2009fluid}
C.-k. Kang, Y.~Baik, L.~Bernal, M.~Ol, and W.~Shyy, ``Fluid dynamics of pitching and plunging airfoils of reynolds number between 1$\times$ 10\^{} 4 and 6$\times$ 10\^{} 4,'' in {\em 47th AIAA aerospace sciences meeting including the new horizons forum and aerospace exposition}, p.~536, 2009.

\bibitem{mcgowan2011investigations}
G.~Z. McGowan, K.~Granlund, M.~V. Ol, A.~Gopalarathnam, and J.~R. Edwards, ``Investigations of lift-based pitch-plunge equivalence for airfoils at low reynolds numbers,'' {\em AIAA journal}, vol.~49, no.~7, pp.~1511--1524, 2011.

\bibitem{baik2012unsteady}
Y.~S. Baik, L.~P. Bernal, K.~Granlund, and M.~V. Ol, ``Unsteady force generation and vortex dynamics of pitching and plunging aerofoils,'' {\em Journal of Fluid Mechanics}, vol.~709, pp.~37--68, 2012.

\bibitem{liu2015unsteady}
T.~Liu, S.~Wang, X.~Zhang, and G.~He, ``Unsteady thin-airfoil theory revisited: application of a simple lift formula,'' {\em Aiaa Journal}, vol.~53, no.~6, pp.~1492--1502, 2015.

\bibitem{cordes2017note}
U.~Cordes, G.~Kampers, T.~Mei{\ss}ner, C.~Tropea, J.~Peinke, and M.~H{\"o}lling, ``Note on the limitations of the theodorsen and sears functions,'' {\em Journal of Fluid Mechanics}, vol.~811, p.~R1, 2017.

\end{thebibliography}
\bibliographystyle{ieeetr}

\end{document}